\documentclass[aps,prl,reprint,superscriptaddress]{revtex4-2}
\usepackage{amsmath, amssymb, braket, dsfont,bm}
\usepackage{subfigure}
\usepackage{graphicx}
\usepackage{rotating, mathtools}
\usepackage[matrix,arrow]{xy}
\usepackage{comment}
\usepackage[pdftex,colorlinks,urlcolor=teal,citecolor=blue,linkcolor=blue]{hyperref}
\usepackage{tikz}
\usepackage{listings}

\usetikzlibrary{calc}
\usetikzlibrary{decorations}
\usetikzlibrary{decorations.markings}
\usetikzlibrary{plotmarks}
\usetikzlibrary{arrows.meta}

\newcommand{\be}{\begin{equation}}
\newcommand{\ee}{\end{equation}}
\newcommand{\bea}{\begin{eqnarray}}
\newcommand{\eea}{\end{eqnarray}}

\newcommand{\Ai}{{\mathbf{A}}_1}
\newcommand{\Ao}{{\mathbf{A}}_0}

\newcommand{\aux}{a} 

\newcommand{\rmi}{\mathrm{i}}

\begin{document}

\title{Lattice non-invertible symmetry from non-commuting transfer matrices}

\author{Eric Vernier}
\affiliation{Université Paris Cité and Sorbonne Université, CNRS, Laboratoire de Probabilités, Statistique et Modélisation, F-75013 Paris, France}
\author{Yuan Miao}
\affiliation{Laboratoire de Physique de l'École Normale Supérieure, ENS, Université PSL, CNRS, Sorbonne Université, Université Paris Cité, Paris, France}
\affiliation{Kavli Institute for the Physics and Mathematics of the Universe (WPI), UTIAS, The University of Tokyo, Kashiwa, Chiba 277-8583, Japan}
\author{Masahito Yamazaki}
\affiliation{Graduate School of Physics, University of Tokyo, Tokyo 113-0033, Japan}
\affiliation{Kavli Institute for the Physics and Mathematics of the Universe (WPI), UTIAS, The University of Tokyo, Kashiwa, Chiba 277-8583, Japan}
\affiliation{Trans-Scale Quantum Science Institute, University of Tokyo, Tokyo 113-0033, Japan}

\begin{abstract}
{\color{black}
Conventional quantum integrability is encoded in a commuting algebra of transfer matrices. By contrast, several models possess additional non-commuting conserved charges with important physical consequences, yet the nature of the corresponding symmetry has remained elusive.  
Focusing on the XXZ spin chain at roots of unity, we show that the non-Abelian analogue of the commuting transfer-matrix algebra is governed by quadratic relations following from a new class of unbalanced Yang--Baxter/RLL relations.
This quadratic algebra is shown to encode precisely the Onsager algebra, for which we construct explicit matrix-product representations of both its generators and its duality defect line. The latter obeys $\mathbb{Z}_N$ Tambara--Yamagami fusion rules, thereby providing a lattice realization of the topological defect lines of the compactified boson conformal field theory. Our results identify non-Abelian transfer-matrix algebras as a microscopic origin for Onsager symmetry and dualities in lattice models.
}
\end{abstract}

\maketitle

\emph{Introduction.}--- Symmetry and integrability are deeply intertwined in the study of quantum many-body systems. On the one hand, quantum integrability is defined by the celebrated Yang--Baxter equation \cite{Yang:1967bm,Baxter:1972hz, baxter1985exactly}, which guarantees the existence of local or quasilocal \cite{Faddeev:1996iy, Prosen:2011veg, Prosen:2013woz} conserved quantities that govern, for instance, late-time relaxation dynamics \cite{Ilievski:2015jhc, Ilievski:2016fdy}. On the other hand, symmetries in modern language \cite{Gaiotto:2014kfa} are represented by topological defect lines (TDL) \cite{Verlinde:1988sn,Frohlich:2006ch,Aasen:2016dop,Aasen_2020,Chang:2018iay}, which are in general non-invertible \cite{Schafer-Nameki:2023jdn,Shao:2023gho} and are described by fusion categories \cite{Tensor_categories}; such non-invertible symmetries constrain the dynamics of the theory \cite{Komargodski:2020mxz} and extend the standard Landau paradigm \cite{Bhardwaj:2023fca,Chen:2025uno} in the classification of phases of matter, for example. Despite these profound individual developments, symmetry and integrability are often discussed separately, and the interplay between the two remains a rich area with much room for further exploration. 

In this Letter, we provide a novel connection between 
symmetry and integrability. 
In short, we introduce a non-Abelian generalization of the Yang–Baxter equation in integrable lattice models whose conserved quantities do not all commute with one another. 
This structure leads to an explicit construction of topological defect lines satisfying non-invertible fusion rules, thereby realizing non-invertible symmetries directly on the lattice.

We discuss the one-dimensional spin-1/2 XXZ quantum chain, with transfer matrices inherited from its two-dimensional classical counterpart, the six-vertex model~\cite{lieb1972twodimensional, baxter1985exactly}. 
We focus on the root-of-unity points, which are dense in the gapless phase of the XXZ model. In this regime, the non-semisimplicity of the underlying quantum group \cite{MR982278,MR1103601} results in an enhanced symmetry through the presence of additional transfer matrices which commute with the Hamiltonian but not necessarily with one another, resulting in exponential degeneracies of the energy spectrum \cite{Fabricius_2001_1, Fabricius_2001_2,baxter2002completeness,MLP_2021}. 
Building on this structure, we derive an unbalanced generalization of the  Yang--Baxter  relation, providing the starting point for the non-Abelian transfer-matrix algebra.

We show that the resulting quadratic transfer-matrix algebra realizes the Onsager algebra~\cite{Onsager, Davies_1990}, an infinite-dimensional Lie algebra originally introduced in the solution of the two-dimensional Ising model and which has recently seen renewed interest in connection with quantum scars~\cite{o2020tunnels,shibata2020onsager} and symmetry-protected topological phases \cite{jones2025pivoting}. While the existence of an Onsager-type symmetry has long been suspected in the periodic XXZ chain at roots of unity~\cite{Deguchi_2001, Korff_2004, Deguchi_2007, Vernier:2018han, Miao_2021}, an explicit construction of its generators has so far remained elusive~\footnote{We remark that a $\mathfrak{sl}_2$ loop algebra symmetry of XXZ spin chain at root of unity was proposed by \cite{Deguchi_2001, Deguchi_2007}. However, The $\mathrm{L}(\mathfrak{sl}_2)$ generators in \cite{Deguchi_2001, Korff_2004, Deguchi_2007} only act on a sector of the entire Hilbert space of the spin chain. On the contrary, the Onsager generators in our paper act on the entire Hilbert space, and commute with the Hamiltonian over the entire Hilbert space, thus a \emph{bona fide} symmetry. One may notice that the Onsager algebra is a fixed-point Lie subalgebra of $\mathrm{L}(\mathfrak{sl}_2)$~\cite{El-Chaar_2012}, but to our knowledge, our Onsager generators seem to be different from the $\mathrm{L}(\mathfrak{sl}_2)$ generators in \cite{Deguchi_2001, Korff_2004, Deguchi_2007}.}.

In its original formulation, where its generators act on a quantum Ising chain, the Onsager algebra is equipped with a duality automorphism identified with the Kramers--Wannier duality \cite{KW_1941_1, KW_1941_2}. The latter has recently attracted renewed interest \cite{Aasen:2016dop, Aasen_2020, Lootens:2021tet, Lootens:2022avn, 
Li:2023ani, Cao:2023doz, Seiberg:2023cdc} as a prototypical example of a non-invertible symmetry. 
For the XXZ spin chain at root of unity, we show that an explicit realization of the duality can be constructed from the non-commuting transfer matrices, thereby yielding a non-local, non-invertible symmetry of the model. 
Finally, we show that this duality operator provides a lattice realization of a Tambara--Yamagami non-invertible symmetry in the compactified-boson CFT describing the continuum limit of the XXZ chain. 
{
Together, our results suggest that non-Abelian transfer-matrix algebras provide the natural language for understanding the interplay between integrability and non-invertible symmetry.}

\emph{The model.}--- We consider the XXZ Hamiltonian acting on a one-dimensional chain of $L$ spins-1/2:
\be 
H_{\rm XXZ} = \sum_{j=1}^L \left(\sigma_j^+ \sigma_{j+1}^- +   \sigma_j^- \sigma_{j+1}^+ + \frac{\Delta}{2} \sigma_j^z \sigma_{j+1}^z \right) \,,
\label{eq:Hamiltonian}
\ee  
where $\sigma^\alpha_j$ are the Pauli matrices acting at position $j$. The anisotropy parameter $\Delta$ is parametrized by a complex number $q$ via $\Delta = (q+q^{-1})/2$.
In this Letter, we will focus on the case where $q$ is a root of unity: setting $\omega \equiv q^2$, we define $N$ to be the smallest integer such that $\omega^N=1$. This implies that $q= e^{\rmi \pi {M}/{N}}$, where $M$ and $N$ are coprimes. 
Twisted boundary conditions are defined as $\sigma_{L+1}^\pm = q^{\mp L} \sigma_1^\pm$, $\sigma_{L+1}^z=\sigma_1^z$. Physically, they correspond to inserting a magnetic field in the one-dimensional ring where the spins live, leading to a non-vanishing phase for the spin exchange interaction.
This simplifies the subsequent presentation; however, our main results also hold for periodic systems, with the two cases equivalent for $L$ a multiple of $2N$.
By a simple unitary change of basis, the twist can be spread evenly over the entire chain: the Hamiltonian, which we simply denote as $H$, then has each $\sigma_j^\pm \sigma_{j+1}^\mp$ multiplied by $q^{\pm 1}$, and genuine periodic boundary conditions. 

\begin{figure}
\begin{tikzpicture}
\newcommand\rad{0.2}
\node at (-1.,0) {\large $T_{\rm 6v} (u)=$};
\draw[line width=1,-<,>=latex] (0,0)-- (4.75,0);
\foreach \x in {0.5,1.25,...,4.5} 
{
\draw[line width=1,>-,>=latex] (\x,-0.5) -- (\x,0.5);
\draw[gray,fill=gray!10,thick]  (\x,0) circle (\rad); \node at (\x,0) {\small  ${R}$};
}
\node at (5,0.) {\scriptsize $a$};
\node at (0.5,-0.7) {\scriptsize $1$};
\node at (1.25,-0.7) {\scriptsize $2$};
\node at (4.25,-0.7) {\scriptsize $L$};

\end{tikzpicture}

\vspace{0.1cm}

\begin{tikzpicture}
\newcommand\shift{1.1}
\draw[line width=1] (0,-0.4) -- (0.,0.4);
\draw[line width=1] (-0.4,0) -- (0.4,0.);

\draw[line width=0.75,-<] (0,-0.3) -- (0.,-0.31);
\draw[line width=0.75,->] (0,0.3) -- (0.,0.31);
\draw[line width=0.75,-<] (-0.3,0) -- (-0.31,0);
\draw[line width=0.75,->] (0.3,0) -- (0.31,0);
\node at (0,-0.75) {$\mathsf{a}$};

\begin{scope}[shift={(\shift,0)}]
\draw[line width=1] (0,-0.4) -- (0.,0.4);
\draw[line width=1] (-0.4,0) -- (0.4,0.);

\draw[line width=0.75,->] (0,-0.3) -- (0.,-0.31);
\draw[line width=0.75,-<] (0,0.3) -- (0.,0.31);
\draw[line width=0.75,->] (-0.3,0) -- (-0.31,0);
\draw[line width=0.75,-<] (0.3,0) -- (0.31,0);
\node at (0,-0.75) {$\mathsf{a}$};
\end{scope}

\begin{scope}[shift={(2*\shift,0)}]
\draw[line width=1] (0,-0.4) -- (0.,0.4);
\draw[line width=1] (-0.4,0) -- (0.4,0.);

\draw[line width=0.75,->] (0,-0.3) -- (0.,-0.31);
\draw[line width=0.75,-<] (0,0.3) -- (0.,0.31);
\draw[line width=0.75,-<] (-0.3,0) -- (-0.31,0);
\draw[line width=0.75,->] (0.3,0) -- (0.31,0);
\node at (0,-0.75) {$\omega \mathsf{b}$};
\end{scope}

\begin{scope}[shift={(3*\shift,0)}]
\draw[line width=1] (0,-0.4) -- (0.,0.4);
\draw[line width=1] (-0.4,0) -- (0.4,0.);

\draw[line width=0.75,-<] (0,-0.3) -- (0.,-0.31);
\draw[line width=0.75,->] (0,0.3) -- (0.,0.31);
\draw[line width=0.75,->] (-0.3,0) -- (-0.31,0);
\draw[line width=0.75,-<] (0.3,0) -- (0.31,0);
\node at (0,-0.75) {$\mathsf{b}$};
\end{scope}

\begin{scope}[shift={(4*\shift,0)}]
\draw[line width=1] (0,-0.4) -- (0.,0.4);
\draw[line width=1] (-0.4,0) -- (0.4,0.);

\draw[line width=0.75,->] (0,-0.3) -- (0.,-0.31);
\draw[line width=0.75,->] (0,0.3) -- (0.,0.31);
\draw[line width=0.75,-<] (-0.3,0) -- (-0.31,0);
\draw[line width=0.75,-<] (0.3,0) -- (0.31,0);
\node at (0,-0.75) {$\mathsf{c}$};
\end{scope}

\begin{scope}[shift={(5*\shift,0)}]
\draw[line width=1] (0,-0.4) -- (0.,0.4);
\draw[line width=1] (-0.4,0) -- (0.4,0.);

\draw[line width=0.75,-<] (0,-0.3) -- (0.,-0.31);
\draw[line width=0.75,-<] (0,0.3) -- (0.,0.31);
\draw[line width=0.75,->] (-0.3,0) -- (-0.31,0);
\draw[line width=0.75,->] (0.3,0) -- (0.31,0);
\node at (0,-0.75) {$\mathsf{c}$};
\end{scope}
\end{tikzpicture}
\caption{
The transfer matrix of the (asymmetric) six-vertex model. (Top): Spatial structure as a matrix product operator; (Bottom): possible configuration of arrows at each vertex. The corresponding weights are given by $\mathsf{a}=e^{2u} \omega -1$, $\mathsf{b}=e^{2u}-1$, $\mathsf{c}= e^u ( \omega-1)$.}
\label{fig:sixvertex}
\end{figure}
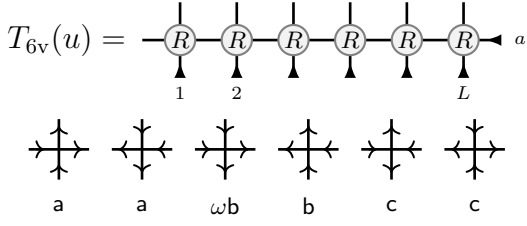

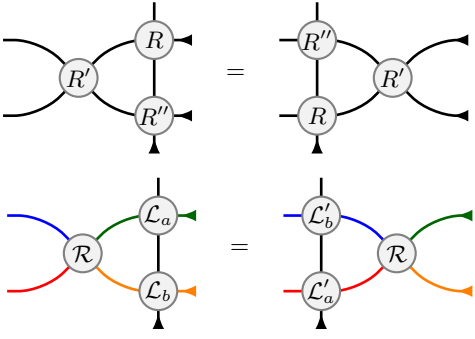
\begin{figure}
\newcommand\rad{0.25}

\begin{tikzpicture}[baseline=(current  bounding  box.center),scale=1.]
\draw[line width=1, rounded  corners=10pt] (0,-0.5) -- (0.5,-0.5) -- (1,0);
\draw[line width=1,-<,>=latex, rounded  corners=10pt] (1,0) -- (1.5,0.5) -- (2.5,0.5);
\draw[line width=1, rounded  corners=10pt] (0,0.5) -- (0.5,0.5) -- (1,0);
\draw[line width=1,-<,>=latex, rounded  corners=10pt] (1,0) -- (1.5,-0.5) -- (2.5,-0.5);
\draw[line width=1,>-,>=latex] (2,-1) -- (2,1);

\draw[gray,fill=gray!10,thick]  (1,0) circle (\rad); \node at (1,0) {$R'$};
\draw[gray,fill=gray!10,thick]  (2,0.5) circle (\rad); \node at (2,0.5) {$R$};
\draw[gray,fill=gray!10,thick]  (2,-0.5) circle (\rad); \node at (2,-0.5) {$R''$};
\end{tikzpicture}
\quad 
=
\quad
\begin{tikzpicture}[baseline=(current  bounding  box.center),scale=1.]
\draw[line width=1, rounded  corners=10pt] (0.5,-0.5) -- (1.5,-0.5) -- (2,0);
\draw[line width=1,-<,>=latex, rounded  corners=10pt] (2,0) -- (2.5,0.5) -- (3.,0.5);
\draw[line width=1, rounded  corners=10pt] (0.5,0.5) -- (1.5,0.5) -- (2,0);
\draw[line width=1,-<,>=latex, rounded  corners=10pt] (2,0) -- (2.5,-0.5) -- (3.,-0.5);
\draw[line width=1,>-,>=latex] (1,-1) -- (1,1);

\draw[gray,fill=gray!10,thick]  (2,0) circle (\rad); \node at (2,0) {$R'$};
\draw[gray,fill=gray!10,thick]  (1,0.5) circle (\rad);\node at (1,0.5) {$R''$};
\draw[gray,fill=gray!10,thick]  (1,-0.5) circle (\rad); \node at (1,-0.5) {$R$};
\end{tikzpicture}
\vspace{0.25cm}

\begin{tikzpicture}[baseline=(current  bounding  box.center),scale=1.]
\draw[line width=1,red, rounded  corners=10pt] (0,-0.5) -- (0.5,-0.5) -- (1,0);
\draw[line width=1,-<,>=latex,green!40!black, rounded  corners=10pt] (1,0) -- (1.5,0.5) -- (2.5,0.5);
\draw[line width=1,blue, rounded  corners=10pt] (0,0.5) -- (0.5,0.5) -- (1,0);
\draw[line width=1,-<,>=latex,orange, rounded  corners=10pt] (1,0) -- (1.5,-0.5) -- (2.5,-0.5);
\draw[line width=1,>-,>=latex] (2,-1) -- (2,1);

\draw[gray,fill=gray!10,thick]  (1,0) circle (\rad); \node at (1,0) {$\cal R$};
\draw[gray,fill=gray!10,thick]  (2,0.5) circle (\rad); \node at (2,0.5) {${\cal L}_a$};
\draw[gray,fill=gray!10,thick]  (2,-0.5) circle (\rad); \node at (2,-0.5) {${\cal L}_b$};
\end{tikzpicture}
\quad 
=
\quad
\begin{tikzpicture}[baseline=(current  bounding  box.center),scale=1.]
\draw[line width=1,red, rounded  corners=10pt] (0.5,-0.5) -- (1.5,-0.5) -- (2,0);
\draw[line width=1,-<,>=latex,green!40!black, rounded  corners=10pt] (2,0) -- (2.5,0.5) -- (3.,0.5);
\draw[line width=1,blue, rounded  corners=10pt] (0.5,0.5) -- (1.5,0.5) -- (2,0);
\draw[line width=1,-<,>=latex,orange, rounded  corners=10pt] (2,0) -- (2.5,-0.5) -- (3.,-0.5);
\draw[line width=1,>-,>=latex] (1,-1) -- (1,1);

\draw[gray,fill=gray!10,thick]  (2,0) circle (\rad); \node at (2,0) {$\cal R$};
\draw[gray,fill=gray!10,thick]  (1,0.5) circle (\rad); \node at (1,0.5) {${\cal L}'_b$};
\draw[gray,fill=gray!10,thick]  (1,-0.5) circle (\rad); \node at (1,-0.5) {${\cal L}'_a$};
\end{tikzpicture}

\caption{Top: 
The Yang--Baxter equation ensuring the commutativity of the six-vertex transfer matrices.
Bottom : 
The unbalanced RLL equation, which encodes the algebra of non-commuting transfer matrices.
}
\label{fig:RLL}
\end{figure}

The model is integrable for any $\Delta$, meaning that $H$ commutes with an extensive number of local charges. Those are generated by the transfer matrices of the six-vertex model, defined as 
\be 
T_{\rm 6v}(u)= \mathrm{Tr}_\aux (R_{\aux 1}(u)\cdots R_{\aux L}(u))  \,,
\ee
where $\aux$ denotes a two-dimensional auxiliary space and the $4\times4$ R-matrices encode the local weights for each vertex configuration, see Fig.~\ref{fig:sixvertex}  (due to the twist, we are here dealing with an asymmetric version of the weights). The transfer matrices $T_{\rm 6v}(u)$ commute with one another for different values of the parameter $u$ as well as with the Hamiltonian. This follows from the Yang--Baxter equation, depicted on the top panel of Fig.~\ref{fig:RLL}:
{
applied repeatedly from one end of the system to the other (the so-called ``train argument''), it exchanges the order of two transfer matrices, implying their commutativity.
}
In addition to the local charges, quasilocal conserved charges can be generated from transfer matrices where the auxiliary space $a$ carries higher-dimensional representations of the quantum group $U_q(\mathfrak{sl}_2)$, which for generic $q$ can be viewed as deformations of the (half)-integer spin representations of $SU(2)$~\cite{Prosen:2013woz}. 
{
Concretely, these transfer matrices take the form of matrix product operators (MPO) 
\be 
T_S(u) = \mathrm{Tr}_a(\mathcal{L}_{a,1}(u) \ldots \mathcal{L}_{a,L}(u)) \,,
\quad
S=\tfrac{1}{2},1,\ldots.
\label{eq:TSMPO}
\ee 
where each L-operator $\mathcal{L}_{a,j}(u)$ acts between a physical spin $j$ and the auxilliary space $a=\mathbb{C}^{2S+1}$. 
}
They commute with $T_{6\mathrm{v}}(u)=T_{1/2}(u)$ as well with one another, as can be derived from variants of the Yang--Baxter equation, the so-called RLL equations, where one (or two) lines carry a higher spin instead of a spin-1/2. As a consequence, for generic $q$, the symmetry of the XXZ chain model can be represented by families of generating functions $\mathcal{X}_S(u)\equiv \partial_u \log T_S(u)$, which generate local (resp. quasilocal) charges for $S=\tfrac{1}{2}$ (resp. $S\geq 1$)~\cite{Ilievski:2015jhc}. 

\begin{table}
\centering
\[
\begin{array}{c|c|c}
&
\text{$q$ generic}
&
\text{$q^N=\pm1$}
\\
\hline
\text{commuting TMs}
& \{T_S\}_{S=\frac12,1,\ldots}
& \{T_S\}_{S=\frac12,1,\ldots}
\\
\text{non-commuting TMs}
& \text{---}
& \bm T_{\rm c},\ \bm T_{\rm n}
\\
\hline
\text{algebra}
& [T,T]=0
& \begin{array}{c}
[T,T]=0\\[0mm]
[\bm T,T]=0\\[0mm]
[\bm T,\bm T]~\text{generically}~\neq0
\end{array}
\end{array}
\]
\caption{
{
Transfer matrices commuting with the Hamiltonian $H$ for generic $q$ and at roots of unity. The root-of-unity case features additional transfer matrices (shown in bold), which give rise to a non-abelian algebra.}}
\label{recaptable}
\end{table}

A distinctive feature of roots of unity, usually referred to as {\it superintegrability}~\cite{Davies_1990}~\footnote{Compared to the superintegrability in models with Yangian symmetry, such as the Calogero--Sutherland model~\cite{Isachenkov:2016gim}, the non-Abelian conserved charges of the XXZ spin chain at roots of unity are \emph{local} or \emph{quasilocal}.}, is the emergence of additional conserved quantities, {with important consequences for nonequilibrium dynamics, transport, scar states, and spectral degeneracies \cite{Prosen:2013woz,ares2023,VernierBertini2023}}. 
These follow from representations with no analogue in $SU(2)$, as a result of the quantum group becoming non-semisimple: for $\omega^N=1$ such representations are $N$-dimensional, and usually referred to as (semi)cyclic or nilpotent \cite{MR1103601,Date_1991}. 
{
The resulting transfer matrices, which depend respectively on 5 and 3 continuous parameters, can be written in the MPO form \eqref{eq:TSMPO}, with auxiliary space $a=\mathbb{C}^N$. 
We denote them by bold symbols $\bm T_{\rm c}(u,v,s,y,y')$ and $\bm T_{\rm n}(u,v)$, reserving ordinary symbols for the conventional transfer matrices.
For convenience, the different families of transfer matrices are summarized in Table~\ref{recaptable}, and we refer to the Supplementary Material for explicit definitions. 
}
The (semi)-cyclic transfer matrices are also famously related to the study of the chiral Potts model \cite{Bazhanov:1989nc,Au-YangMcCoyBarryPerkTangYan1987,BaxterPerkAu-Yang1988}, being identified as the column-to-column transfer matrices of the $\tau_2$ model~\cite{Baxter:1990mft,baxter2004transfer,Roan_2009}, and can be factorized as $\bm T_{\rm c}(u,v,s,y,y')=e^{y S^z} \bm T_{\rm c}(u,v,s) e^{y' S^z}$, where $S^z= \frac{1}{2}\sum_{j=1}^L \sigma_j^z$ is the conserved magnetization.  

The matrices $\bm T_{\rm c}$ and $\bm T_{\rm n}$ commute with $T_{6 \mathrm{v}}(u)$ (and therefore with $H$),
\be 
[\bm T_{\rm c}(u,v,s), T_{\rm 6v} (x)] = [\bm T_{\rm n}(u,v), T_{\rm 6v} (x)] =0  \,,
\label{eq:6vtau2commute}
\ee 
as well as with the transfer matrices $T_S$ built from (half)-integer spin auxiliary representations. However, they generally do not commute among themselves, and generate a non-Abelian symmetry that manifests itself as large degenerate multiplets in the energy spectrum. 
{
Although the transfer matrices introduced above are generally non-commuting, they admit several important commuting subfamilies. In particular, the $\bm T_{\rm n}$ commute among themselves. Moreover, when all parameters except $u$ vanish, both families reduce to the conventional commuting transfer matrix,
}
\be 
\bm T_{\rm c}(u,0,0)= \bm T_{\rm n}(u,0) = T_{\frac{N-1}{2}}(u)\equiv T(u) \,.
\ee 

\begin{figure}
\newcommand\rad{0.25}

\centering

\begin{tikzpicture}[baseline=(current  bounding  box.center),scale=1.]
\draw[line width=1,red, rounded  corners=10pt] (0,-0.5) -- (0.5,-0.5) -- (1,0);
\draw[line width=1,-<,>=latex,green!40!black, rounded  corners=10pt] (1,0) -- (1.5,0.5) -- (4.5,0.5);
\draw[line width=1,blue, rounded  corners=10pt] (0,0.5) -- (0.5,0.5) -- (1,0);
\draw[line width=1,-<,>=latex,orange, rounded  corners=10pt] (1,0) -- (1.5,-0.5) -- (4.5,-0.5);
\foreach \x in {1.75,2.5,...,4} 
{
\draw[line width=1,>-,>=latex] (\x,-1) -- (\x,1);
\draw[gray,fill=gray!10,thick]  (\x,0.5) circle (\rad); \node at (\x,0.5) {\small  ${\cal L}_a$};
\draw[gray,fill=gray!10,thick]  (\x,-0.5) circle (\rad); \node at (\x,-0.5) {\small  ${\cal L}_b$};
}
\draw[gray,fill=gray!10,thick]  (1,0) circle (\rad); \node at (1,0) {\footnotesize ${\cal R}$};
\end{tikzpicture}
\quad 
=
\quad
\begin{tikzpicture}[baseline=(current  bounding  box.center),scale=1.]
\draw[line width=1,red, rounded  corners=10pt] (-1.5,-0.5) -- (1.5,-0.5) -- (2,0);
\draw[line width=1,-<,>=latex,green!40!black, rounded  corners=10pt] (2,0) -- (2.5,0.5) -- (3.,0.5);
\draw[line width=1,blue, rounded  corners=10pt] (-1.5,0.5) -- (1.5,0.5) -- (2,0);
\draw[line width=1,-<,>=latex,orange, rounded  corners=10pt] (2,0) -- (2.5,-0.5) -- (3.,-0.5);

\foreach \x in {1.25,0.5,...,-1} 
{
\draw[line width=1,>-,>=latex] (\x,-1) -- (\x,1);
\draw[gray,fill=gray!10,thick]  (\x,0.5) circle (\rad); \node at (\x,0.5) {\small  ${\cal L}'_b$};
\draw[gray,fill=gray!10,thick]  (\x,-0.5) circle (\rad); \node at (\x,-0.5) {\small  ${\cal L}'_a$};
}
\draw[gray,fill=gray!10,thick]  (2,0) circle (\rad); \node at (2,0) {\small $\cal R$};
\end{tikzpicture}

\caption{
The train argument is used to derive quadratic relations between non-commuting transfer matrices.
}
\label{fig:nonabelianRTT}
\end{figure}
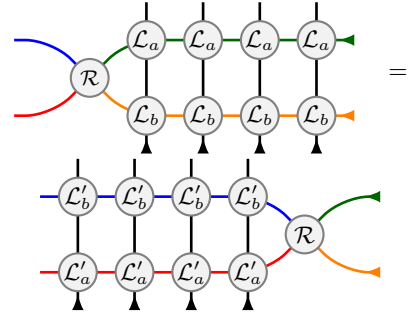

\emph{Unbalanced Yang-Baxter equation and Onsager algebra symmetry.}---
{
While the above construction of transfer matrices is standard, the non-Abelian analogue of the commuting algebra of conserved charges underlying conventional integrability has remained elusive. As the first main result of this Letter, we show that the non-commuting conserved charges generate precisely the Onsager algebra, providing a non-Abelian extension of the Abelian algebra generated by the families $T_S(u)$.  More fundamentally, we show that this Onsager algebra is encoded in a quadratic algebra of transfer matrices, playing for non-Abelian integrability the role that mutual commutativity plays in conventional integrability. 
}
Just as mutual commutativity follows from repeated applications of the Yang--Baxter equation through the train argument, we derive these quadratic relations from a new unbalanced version of the Yang--Baxter equation (the ``unbalanced RLL relation''), which takes the general form
\be 
 \mathcal{R}_{a,b} \mathcal{L}_a \mathcal{L}_b = \mathcal{L}'_b \mathcal{L}'_a \mathcal{R}_{a,b} \,,
\label{eq:nonabelianRLL} 
\ee 
see Fig.~\ref{fig:RLL}. 
Here each side acts on the tensor product of two $N$-dimensional auxiliary spaces, labeled as $a$ and $b$ (colored lines on the figure), and one physical spin-1/2, (black line). 
We call an RLL relation unbalanced when the outgoing L-operators carry transformed parameters, so that the train argument yields a nontrivial product identity {of the form $\bm T_a \bm T_b = \bm T'_b \bm T'_a$}, rather than commutativity  (see Fig.~\ref{fig:nonabelianRTT} for an illustration).  
A list of all such equations relevant for the present Letter is provided in the SM.

{
We now describe the key steps which allow to prove the quadratic relations encoding the Onsager algebra symmetry} (details of the calculations are given in the SM).
First, we prove that the non-commuting transfer matrices introduced above can be factorized as 
\bea 
T(0)^2~\bm T_{\rm c}(u,v,s) &=& \bm T_+(\tfrac{u+ v}{2})~ \bm T_{\rm c}(0,0,s) ~ \bm T_{-}(\tfrac{u- v}{2}) \,,
\label{eq:Ttau2factorization}
\\
T(0) \bm T_{\rm n}(u,w) &=& \bm T_{\rm L}(\tfrac{u+w}{2}) \bm T_{\rm R}(\tfrac{u-w}{2})
\label{eq:Tnilpfactorization}
\,,
\eea  
where $\bm T_\pm (u) \equiv \bm T_\mathrm{c}(u,\pm u,0)$, $\bm T_{\rm L/R}(u) \equiv \bm T_\mathrm{n}(u,\pm u)$~\footnote{We remark that a similar factorization has appeared in the construction of the Q-operator for the XXZ spin chain~\cite{Bazhanov:1989nc, MLP_2021, Weston:2024dml}. We anticipate the close relation between $T_\pm (u)$, $T_{\rm L/R}(u)$, and the Q-operator, though a rigorous proof such as the one in Ref.~\cite{Weston:2024dml} is out of the scope of the current Letter.}. The factorization \eqref{eq:Ttau2factorization} follows in two steps from the identities $T(0) \bm T_{\rm c}(u,v,s) = \bm T_{\rm c}(\tfrac{u+v}{2},\tfrac{u+v}{2},s) \bm T_-(\tfrac{u-v}{2}) = \bm T_+(\tfrac{u+v}{2}) \bm T_{\rm c}(\tfrac{u-v}{2},\tfrac{u-v}{2},s)$, which both derive from unbalanced RLL relations through the train argument. 
The second factorization \eqref{eq:Tnilpfactorization} is dealt with similarly. 
{
Noticing that the identity $\bm T_{\rm c}(0,0,s)\bm T_{\rm c}(0,0,t) =T(0) \bm T_{\rm c}(0,0,s+t)$ also holds, consequence of another unbalanced RLL equation, we can further write the exponential form
\be
\bm T_{\rm c}(0,0,s)=T(0)e^{\frac{s N}{2} \Ai}  \,, ~
\Ai\equiv\frac{2}{N}\partial_s \log \bm T_{\rm c}(0,0,s)|_{s=0} \,.
\label{eq:exponentialformTA1}
\ee 
}
%

Our first contact with the Onsager algebra is made by observing that $\Ai$, together with $\Ao \equiv \frac{2}{N}\sum_{j=1}^L \sigma_j^z$, satisfy the Dolan--Grady relations \cite{DolanGrady}
\bea 
\label{DG1}
{}\Big[ \Ao , \big[ \Ao , [ \Ao , \Ai ] \big] \Big] &=& 16 [ \Ao , \Ai ] \,,  \\
{}\Big[ \Ai , \big[ \Ai , [ \Ai , \Ao ] \big] \Big] &=& 16 [ \Ai , \Ao ] \,.
\label{DG2}
\eea
It is known that two operators obeying \eqref{DG1}, \eqref{DG2} generate a representation of the Onsager algebra \cite{Davies_1990}. The latter is usually formulated in terms of two infinite families of generators $\mathbf{A}_m$, $\mathbf{G}_l$, $m,l\in\mathbb{Z}$, subject to the commutation relations
$[\mathbf{A}_l, \mathbf{A}_m]=4\mathbf{G}_{l-m}$, $[\mathbf{G}_l,\mathbf{A}_m]=2\mathbf{A}_{l+m}-2\mathbf{A}_{m-l}$, and $[\mathbf{G}_l, \mathbf{G}_m]=0$.
{
The proof of the Dolan--Grady relations \eqref{DG1}, \eqref{DG2} exploits the exponential form \eqref{eq:exponentialformTA1} to recast products of exponentials of $\Ao$ and $\Ai$ as product of non-commuting transfer matrices, which can then be simplified using elementary linear algebra techniques. Comparing the result with the Baker-Campbell-Hausdorff formula yields the desired relations.
}

{
Having identified the Onsager algebra, we then show that the entire non-Abelian symmetry can be constructed from its generators.
}
To this end, we write the remaining operators entering the factorization \eqref{eq:Ttau2factorization}, \eqref{eq:Tnilpfactorization} can be written as 
\bea
 \bm T_\pm(u) &=& T(0) \exp\left(\pm\frac{ 1}{2}\mathcal{G}(u)+ \frac{\mathrm{i}}{2} \mathcal{H}(u)\right)\, ,
 \label{eq:Tpmexp}
 \\
 \bm T_{\rm L/R}(u) &=& T(0) \exp\left(\pm\frac{ 1}{2}\mathcal{Q}(u)+ \frac{\mathrm{i}}{2} \mathcal{H}(u)\right) \,,
 \label{eq:TLRexp}
\eea
where $\mathcal{H}(u)$ is the generating function of abelian quasi-local charges associated with $T(u)$, 
\be 
T(u) = T(0) \exp{\mathrm{i} \mathcal{H}(u)} \,, \qquad  \mathrm{i} \partial_u \mathcal{H}(u) =\mathcal{X}_{\frac{N-1}{2}}(u) \,,
\label{eq:Tu}
\ee  
while $\mathcal{G}(u)$ and $\mathcal{Q}(u)$ each admit a series expansions in terms of a mutually commuting family of Onsager generators, namely
\bea  
\mathcal{G}(u) &=& \sum_{m\geq 1} \frac{\tau(u)^m}{m} \mathbf{G}_m \,,
\quad \tau(u)\equiv\tanh(N u)
\label{eq:Gseries}
\\
\mathcal{Q}(u) &=& -\sum_{m\geq 1} \frac{\tau(u)^m}{m} (\mathbf{A}_m+\mathbf{A}_{-m})  -  \tau(u)^2 \Ao\,. 
\eea 
%
When the system size is commensurate with $N$ (when $L$ is a multiple of $2N$), $T(0)$ is invertible and the factorizations \eqref{eq:Ttau2factorization}, \eqref{eq:Tnilpfactorization} can be written under a more compact form, directly expressing $\bm T_{\rm c}$ as ($T(0)$ times) a product of exponentials.

\emph{Duality.}--- 
{
The identification of the Onsager symmetry naturally raises the question of its duality automorphism.}
At the free-fermionic point ($N=2$), the Onsager generators boil down to their original Ising representation,  $\Ao=\sum_j{\sigma_j^z}$ and $\Ai=\sum_j{\sigma_j^x\sigma_{j+1}^x }$, and get exchanged by the Kramers--Wannier duality~\cite{KW_1941_1, KW_1941_2, Miao_2022, Seiberg:2023cdc, Korepin_2026}.
It is known how to implement the latter on the lattice as a non-invertible MPO $D_{\sigma}$, verifying
\begin{equation}
    D_{\sigma} \sigma^z_n = \sigma^x_n \sigma^x_{n+1} D_{\sigma} \; , \,\, D_{\sigma} \sigma^x_n \sigma^x_{n+1} = \sigma^z_{n+1} D_{\sigma} \; 
\end{equation}
The duality defect, together with the spin flip defect $D_{\psi}=\prod_{j=1}^L \sigma_j^z$, obey the well-known fusion rules~\cite{Aasen:2016dop} 
\begin{equation}
\begin{split}
&    D_{\sigma} \cdot D_{\sigma} = \left( 1 +D_{\psi} \right)\cdot \mathbb{T}\; ,
    \\
& D_{\sigma} \cdot D_{\psi} = D_{\psi} \cdot D_{\sigma} = D_{\rm \sigma} 
\end{split}
\label{TYZ2}
\end{equation}
which, modulo the  lattice translation operator $\mathbb{T} = \prod_{n=1}^{L-1} \mathcal{P}_{n,n+1}$, are known as those of the Tambara--Yamagami fusion category $\mathrm{TY}(\mathbb{Z}_2)$ \cite{Tambara:1998vmj}. Meanwhile, it is straightforward to check that $D_{\sigma}$ and $D_{\psi}$ are symmetries of $H$ at $N=2$, corresponding to the twisted XX (Dzyaloshinskii--Moriya) Hamiltonian~\cite{Miao_2022, Pace:2024oys}.

For $N>2$, the Onsager generators have a more complicated expression, reflecting their quasilocal nature. However, from the structure of the algebra, it is natural to consider the following automorphism: $\mathbf{A}_n \to \pm \mathbf{A}_{1-n} \, , \mathbf{G}_n \to \mp \mathbf{G}_{1-n}$, with $n \in \mathbb{Z}$, which keeps the algebraic structure intact. As the second main result of this Letter, we show that  operators $D_{\pm}$ that implement this duality can be constructed as a certain limit of $\bm T_{\rm c}$.
In addition, by checking their fusion algebra we show that they provide a lattice realization of the $\mathrm{TY}(\mathbb{Z}_N)$ topological defect lines of the $c=1$ free compactified boson CFT at rational values of radius squared $R^2 \propto {1}/{N}$~\cite{Wang_2024}. 

Define the following Matrix Product Operators 
\begin{align}
    & D_\pm \equiv 
      \lim_{u\to \mp \infty}
    e^{\frac{\rmi \pi}{2N}   \left(L\mp 2S^z \right)}
    e^{\pm u\,L}  \bm T_{\rm c}(u,u,0) \; .
\end{align}
Using unbalanced RLL relations, we show in the SM that these satisfy
\begin{equation}
    D_{\pm} \mathbf{A}_{0} = \pm \mathbf{A}_{1} D_{\pm} \; , \,\,\,\, D_{\pm} \mathbf{A}_{1} = \pm \mathbf{A}_{0} D_{\pm} \; .
    \label{eq:DpmA0A1}
\end{equation}
Since the entire Onsager algebra is generated by iterated commutators of $\Ao$ and $\Ai$, the relations \eqref{eq:DpmA0A1} suffice to prove that $D_\pm$ realize a duality of the algebra.
Using further MPO manipulations, we show that 
\begin{equation}
\begin{split} &
D_\pm \cdot D_\pm =     \left( \sum_{k=0}^{N-1}  {\mathsf{Z}_{\pm}}^k \right)  T(0)  
\\ 
&
    \mathsf{Z}_{\pm} \cdot D_{\pm} = D_{\pm} \cdot \mathsf{Z}_{\pm} = D_{\pm} \;, 
\end{split}
\label{eq:TY1}
\end{equation}
where the operators   $\mathsf{Z}_{\pm} =e^{\frac{\rmi \pi}{N}  \left(L\mp 2S^z \right)} $ obey 
\begin{equation}
    \mathsf{Z}_{\pm}^N = 1 \; , \quad \mathsf{Z}_{\pm}^m \mathsf{Z}_{\pm}^n = \mathsf{Z}_{\pm}^{m+n} \; ,
    \label{eq:TY2}
\end{equation}
realizing a $\mathbb{Z}_N$ subgroup of the $U(1)$ symmetry generated by the conserved magnetization. 
The relations \eqref{eq:TY1}, \eqref{eq:TY2} provide a lattice realization of the fusion category $\mathrm{TY}(\mathbb{Z}_N)$, extending the Ising fusion rules \eqref{TYZ2} to $N>2$. When $L$ is commensurate with $N$, $T(0)$ is invertible and plays an analogous role to the translation operator in \eqref{TYZ2}. 
{
The commutation relation~\eqref{eq:6vtau2commute} implies that the duality operators $D_\pm$  commute with the Hamiltonian. They thus provide explicit lattice realizations of non-invertible symmetry operators for the XXZ chain at roots of unity.
}
The fact that there are two copies of the $\mathrm{TY}(\mathbb{Z}_N)$ relations follows from the existence of an additional $\mathbb{Z}_2$ symmetry, combination of a global spin flip~($\pi$-rotation about the $x$ axis) and of time reversal, which exchanges $D_\pm$.

For anisotropy parameter $-1< \Delta< 1$, the XXZ spin chain is gapless and flows in the IR to the compactified boson CFT with radius $R$, given by the relation $2R^2 = \pi/\gamma$ in the notation of Ref.~\cite{Ginsparg:1988ui}, where $q=e^{\mathrm{i} \gamma}$. For instance, $R=1$ for $\gamma = {\pi}/{2}$ (free fermion point) and $R = \sqrt{{3}/{2}}$ for $\gamma = {\pi}/{3}$. Therefore, the root-of-unity values of $q$ correspond to rational values of $R^2$, which are known to correspond to rational CFTs and have a wealth of TDLs implementing non-invertible symmetries~\cite{Frohlich:2006ch}. The existence of the $\mathrm{TY}(\mathbb{Z}_N)$ TDLs for $R = \sqrt{{N}/{2}}$ (corresponding to $\gamma = \pi/N$) can be understood with help of T-duality~\cite{Ginsparg:1988ui} (which is an invertible operation in CFT). 
We start with the CFT with $R=\sqrt{{N}/{2}}$, and perform the gauging of the discrete symmetry $\mathbb{Z}_N$, a subgroup of the $U(1)$ momentum symmetry (which is the IR limit of the magnetization $U(1)$ symmetry of the lattice model). The resulting CFT has radius $R' = {R}/{N} = {1}/{\sqrt{2N}}$. The ``domain wall'' between the two theories is denoted as $\mathcal{D}$, satisfying the fusion algebra of the non-Abelian object of $\mathrm{TY}(\mathbb{Z}_N)$ category. The new CFT with radius $R' = {1}/{\sqrt{2N}}$ coincides with the original CFT after T-duality $\mathcal{T} $, which shows the equivalence between the CFT with radius $R'$ and the CFT with radius $R'' = {1}/{(2 R')} = \sqrt{{N}/{2}} = R$. Therefore, for the CFT with radius $R = \sqrt{{N}/{2}}$, the operator $\widetilde{\mathcal{D}} = \mathcal{T} \circ \mathcal{D}$ becomes a symmetry satisfying
\begin{equation}
    \widetilde{\mathcal{D}} \circ \widetilde{\mathcal{D}} = \sum_{n=0}^{N-1} \mathcal{Z}^n \; ,
\end{equation}
where $\mathcal{Z}$ is the generator of the $\mathbb{Z}_N$ symmetry.
Comparing with our findings, we conclude that the duality operator $D_{\pm}$ can be regarded as a lattice realization of the symmetry operator $\widetilde{\mathcal{D}} $ in the $c=1$ free compactified boson CFT, sharing the same fusion rules as the non-Abelian object of $\mathrm{TY}(\mathbb{Z}_N)$ category.

\emph{Outlook.}---
We have studied the interplay between integrability and symmetry in models with non-Abelian conserved charges, focusing on the spin-1/2 XXZ chain at roots of unity. Using unbalanced Yang--Baxter/RLL relations, we identified the symmetry generated by (typically quasilocal) conserved charges as the Onsager algebra. Since this algebra arises from a seemingly unrelated corner of exactly solvable models, our results suggest that Onsager-type structures may be a more general feature of integrable systems with enhanced degeneracies~\footnote{This connection appears unrelated to the appearance of the $q$-Onsager algebra in open XXZ chains \cite{baseilhac2005deformed}.}. Extending this framework to higher-rank, higher-spin, and other root-of-unity integrable models remains an open direction. In particular, the relation to $\mathcal{U}_q(\mathfrak{sl}_2)$ symmetry with open boundaries \cite{Pasquier:1989kd, gainutdinov2015counting, gainutdinov2016algebraic} and to affine Temperley--Lieb representations~\cite{Pinet:2022rrc, Hu:2026eap} remains to be understood.

{We constructed the duality operators $D_\pm$ as MPOs realizing the $\mathrm{TY}(\mathbb{Z}_N)$ fusion algebra on the lattice. However, the fusion algebra does not fix the underlying fusion category, since the associator (F-symbols) remains undetermined. Tensor-network methods~\cite{Lootens:2022avn, Rubio:2022nrg} provide a route to extract this data and reconstruct the full fusion category. We leave a detailed analysis of its relation to integrability for future work.}

A central question is whether every non-invertible symmetry (i.e., fusion algebra) admits a lattice realization as a descendant of a non-Abelian integrable structure. A geometric viewpoint suggests that many fusion categories, particularly those associated with quantum groups at roots of unity~\cite{ReshetikhinTuraev}, arise from three-dimensional Chern--Simons theory~\cite{Witten:1988hf}. This structure can be ``Baxterized'' via a field-theoretic version of T-duality~\cite{Yamazaki:2019prm}, leading to the topological-holomorphic four-dimensional Chern--Simons theory, which unifies several aspects of integrability~\cite{Costello:2017dso,Costello:2018gyb,Costello:2019tri}. While most discussions focus on generic $q$, Ref.~\cite{Moosavian:2025nyv} studied the chiral Potts model at roots of unity and its connections to hyperbolic monopoles~\cite{AtiyahMurray1995}, higher-dimensional holomorphic Chern--Simons theories~\cite{Costello202004,BittlestonSkinner202011}, and string theory embeddings~\cite{CostelloYagi201810,AshwinkumarTanZhao201806}. Extending these ideas to a unified framework of higher-dimensional quantum field theory and string theory is a compelling direction for future research.

\noindent\emph{Acknowledgements.}---We are grateful to Paul Fendley and Robert Weston for discussions and continuous encouragement. E.V.~thanks Azat Gainutdinov for discussions on a related project. Y.M. is grateful to Murray Batchelor, Sasha Garbali, Michio Jimbo, Atsuo Kuniba, and Paul Terwilliger for helpful discussions. Y.M.~thanks the University of Melbourne for the hospitality. The work of E.V. was supported by the
ANR under grant ANR-24-CE40-7252. The work of Y.M.~and M.Y.~is supported by the World Premier International Research Center Initiative (WPI), MEXT, Japan.
M.Y.~is also supported in part by the JSPS Grant-in-Aid for Scientific Research (Grant No.~23K25865), and by JST, Japan (CREST Grant No.~JPMJCR26XA, Moonshot R\&D Grant No.~JPMJMS2061).

\let\oldaddcontentsline\addcontentsline
\renewcommand{\addcontentsline}[3]{}
\bibliography{ref}

\let\addcontentsline\oldaddcontentsline

\onecolumngrid
\newpage 

\appendix
\setcounter{equation}{0}
\setcounter{figure}{0}
\renewcommand{\thetable}{S\arabic{table}}
\renewcommand{\theequation}{S\thesection.\arabic{equation}}
\renewcommand{\thefigure}{S\arabic{figure}}
\setcounter{secnumdepth}{2}

\begin{center}
	{\Large \bf Supplemental Material}
\end{center}

\tableofcontents

\section{Reminders on the asymmetric 6-vertex model}
\label{subapp:a6vtau2}

\subsection{Transfer matrix and Hamiltonian}

The R-matrix of the asymmetric 6-vertex model is given as
\begin{equation}
    R_{1,2} (u) = \begin{pmatrix}
        \mathsf{a} & {\color{gray} 0 } & {\color{gray} 0 } & {\color{gray} 0 } \\
        {\color{gray} 0 } & \omega \mathsf{b} & \mathsf{c} & {\color{gray} 0 } \\
        {\color{gray} 0 } & \mathsf{c} & \mathsf{b} & {\color{gray} 0 } \\
        {\color{gray} 0 } & {\color{gray} 0 } & {\color{gray} 0 } & \mathsf{a}
    \end{pmatrix} \; , 
\end{equation}
where, as in the main text,  $\mathsf{a}=e^{2u} \omega -1$, $\mathsf{b}=e^{2u}-1$, $\mathsf{c}= e^u ( \omega-1)$.
It is straightforward to check the Yang--Baxter equation
\begin{equation}
    R_{1,2}(u-v) R_{1,3} (u) R_{2,3} (v) = R_{2,3} (v) R_{1,3} (u) R_{1,2} (u-v) \; ,
\end{equation}
which implies the commutativity of the asymmetric 6-vertex transfer matrices
\begin{equation}
    \left[ T_{\rm 6v} (u) , T_{\rm 6v} (v) \right] = 0 \; , \quad \forall u,v \in \mathbb{C} \; .
\end{equation}
The logarithmic derivative of the transfer matrix yields the Hamiltonian used in the main text,
\begin{equation}
  \, \left. \partial_u \log T_{\rm 6v} (u) \right|_{u=0} = \frac{2}{  q-q^{-1}} H + \frac{L(3q^2 -1)}{2(q^2-1)} \; , 
  \qquad
  H=\sum_{j=1}^L \left(q \sigma^+_j \sigma^-_{j+1} + q^{-1} \sigma^-_j \sigma^+_{j+1}  + \frac{q+q^{-1}}{4} \sigma^z_j \sigma^z_{j+1} \right) \; ,
\end{equation}
unitarily related to the usual XXZ Hamiltonian with twisted boundaries, 
\begin{equation} 
H^\prime = U H U^{-1}  =  \sum_{j=1}^L \left(q^{L\delta_{j,L}} \sigma^+_j \sigma^-_{j+1} + q^{-L \delta_{j,L}} \sigma^-_j \sigma^+_{j+1}  + \frac{q+q^{-1}}{4} \sigma^z_j \sigma^z_{j+1} \right) \,, \qquad U=\prod_{j=1}^L q^{\frac{j}{2} \sigma_j^z} \; ,
\end{equation}
coinciding with eq.~\eqref{eq:Hamiltonian} in the main text. 
The commutation of 6-vertex transfer matrices together with the definition of the Hamiltonian imply $\left[ H , T_{\rm 6v} (u) \right]=0$ for all $u$. 
The Hamiltonian $H$ possesses the global $U(1)$ symmetry associated with the conservation of the magnetization
\begin{equation}
    S^z = \frac{1}{2} \sum_{j=1}^L \sigma^z_j \; ,
\end{equation}
and a $\mathbb{Z}_2$ symmetry that is the combination of the spin-flip and the time-reversal symmetry,
\begin{equation}
    H = \left( \prod_{j=1}^L \sigma^x_j \cdot H \cdot \prod_{j=1}^L \sigma^x_j \right)^\ast \; .
\end{equation}
Together, they form a representation of a global $O(2) = U(1) \rtimes \mathbb{Z}_2$. The $\mathbb{Z}_2$ symmetry will play a role when identifying the duality operators in the lattice setup.

\subsection{Auxiliary transfer matrices at root of unity}

At a root of unity ($q^N=\pm 1$), additional transfer matrices can be defined, related to the (semi)cyclic or nilpotent representations of the quantum group $U_q(sl_2)$.  
The cyclic transfer matrices $\bm T_{\rm c}(u,v,s,y,y')=e^{y S^z} \bm T_{\rm c}(u,v,s) e^{y' S^z}$, also known as the column-to-column transfer matrices of the $\tau_2$ model, involve an auxiliary space $a=\mathbb{C}^N$, on which operators $X_a$ and $Z_a$ act, satisfying the algebra $Z_a X_a = \omega X_a Z_a$, $Z_a^N = X_a^N = 1$. They are expressed in matrix-product form 
\begin{align}
& \bm T_{\rm c}(u,v,s)  = \mathrm{Tr}_{\aux}\left( \mathcal{L}_{\aux,1}(u,v,s) \cdots \mathcal{L}_{\aux,L}(u,v,s)  \right) 
\, ,
\label{eq:Tc}
\\ 
& \mathcal{L}_{\aux,j}
= \left(
\begin{array}{cc}
e^{s-u} + e^{u-s} Z_a &  \omega (e^{s-v} - e^{v-s} Z_a)X_a^{-1} \\
 X_a(e^{v+s} -  e^{-v-s} Z_a) & \omega e^{u+s} + e^{-u-s} Z_a
\end{array} 
\right)\,,
\nonumber
\end{align}
where the $\cal{L}_{\aux,j}$ are known as L--operators, or Lax operators.
For concreteness, we may work in the basis 
\be 
Z_a |n\rangle = \omega^n |n\rangle  \,, \qquad 
X_a^{\pm 1} |n\rangle = |n\pm 1 \mod N \rangle \,, \qquad n=1,\ldots N
\label{eq:Zbasis} \,,
\ee 
The asymmetric 6-vertex R matrix and the $\tau_2$ L--operator satisfy the RLL relation,
\begin{equation}
    \mathcal{L}_{\aux,1} (u_1 - u_2,v,s) \mathcal{L}_{\aux,2} (u_1,v,s) R_{1,2} (u_2)  = R_{1,2} (u_2) \mathcal{L}_{\aux,2} (u_1,v,s) \mathcal{L}_{\aux,1} (u_1-u_2,v,s) \; .
\end{equation}
which implies that the asymmetric 6-vertex transfer matrix, and therefore the Hamiltonian $H$, commutes with the $\tau_2$ transfer matrix
\begin{equation}
    \left[ T_{\rm 6v} (x) , \bm T_{\rm c} (u,v,s) \right]  =   \left[H, \bm T_{\rm c} (u,v,s) \right] = 0 \; , \quad x,u,v,s \in \mathbb{C} \; .
\end{equation}

Transfer matrices based on the nilpotent quantum group representations  (sometimes referred to as ``complex spin'' representations) can conveniently be expressed by introducing in the $a=\mathbb{C}^N$ auxiliary space raising and lowering nilpotent operators $S_a^\pm$ such that $X_a^{\pm 1} = S_a^\pm + (S_a^\mp)^{N-1}$. In the basis \eqref{eq:Zbasis}, these act as $S_a^+ |n\rangle = \delta_{n<N}|n+1\rangle$ and $S_a^- |n\rangle = \delta_{n>1}|n-1\rangle$. 
The transfer matrices are defined as 
\begin{align}
& \bm T_{\rm n}(u,w)  = \mathrm{Tr}_{a}\left( \mathcal{L}_{a,1}(u,w) \ldots \mathcal{L}_{a,L}(u,w)  \right) 
\label{eq:Tn}
\\ 
&
 \mathcal{L}_{a,j}(u,w)
 =  \left(
\begin{array}{cc}
e^{-u-w} + e^{u+w} Z_a &  \omega (e^{-2w} - e^{2w} Z_a)S_a^-  \\
 S_a^+ (1 - Z_a) & \omega e^{u-w} + e^{-u+w} Z_a
\end{array} 
\right)_j 
\,.
\nonumber
\end{align}
The asymmetric 6-vertex R matrix and L--operators $\mathcal{L}_{a,j}(u,w)$ satisfy the RLL relation, 
\begin{equation}
    \mathcal{L}_{\aux,1} (u_1 - u_2,w) \mathcal{L}_{\aux,2} (u_1,w) R_{1,2} (u_2)  = R_{1,2} (u_2) \mathcal{L}_{\aux,2} (u_1,w) \mathcal{L}_{\aux,1} (u_1-u_2,w) \; ,
\end{equation}
which implies the commutativity of transfer matrices
\begin{equation}
    \left[ T_{\rm 6v} (x) , \bm T_{\rm n} (u,w) \right]  =   \left[H, \bm T_{\rm n} (u,w) \right] = 0 \; , \quad x,u,w \in \mathbb{C} \; .
\end{equation}

\noindent{\bf Remark.} The RLL relations considered above also ensure the commutativity between two inhomogeneous transfer matrices. Define the inhomogeneous asymmetric 6-vertex transfer matrix as
\begin{equation}
    T_{\rm 6v}^{\rm inh} (u,\{\xi_i \}_{i=1}^L) = \mathrm{Tr}_a \left( R_{a1} (u-\xi_1) R_{a2} (u-\xi_2) \cdots R_{aL} (u-\xi_L) \right) \; ,
\end{equation}
and analogously the inhomogeneous $\tau_2$ and nilpotent transfer matrices
\begin{align}
    \bm T_{\rm c}^{\rm inh} (u,v,s,\{\xi_i \}_{i=1}^L) &= \mathrm{Tr}_a \left( \mathcal{L}_{a 1} (u-\xi_1,v,s) \mathcal{L}_{a 2} (u-\xi_2,v,s) 
    \ldots \mathcal{L}_{a L} (u-\xi_L,v,s) \right) \; .
    \\
 \bm T_{\rm n}^{\rm inh} (u,w,\{\xi_i \}_{i=1}^L) &= \mathrm{Tr}_a \left( \mathcal{L}_{a 1} (u-\xi_1,w) \mathcal{L}_{a 2} (u-\xi_2,w) 
 \ldots \mathcal{L}_{a L} (u-\xi_L,w) \right) \; .
\end{align}
The RLL relation implies that
\begin{equation}
    \left[ T_{\rm 6v}^{\rm inh} (x,\{\xi_i \}_{i=1}^L) , \bm T_{\rm c}^{\rm inh} (u,v,s,\{\xi_i \}_{i=1}^L) \right] =
       \left[ T_{\rm 6v}^{\rm inh} (x,\{\xi_i \}_{i=1}^L) , \bm T_{\rm n}^{\rm inh} (u,w,\{\xi_i \}_{i=1}^L) \right] 
    =
    0 , \quad x,u,v,s,w \in \mathbb{C} \; ,
\end{equation}
with fixed inhomogeneities $\{\xi_i \}_{i=1}^L$. It allows in particular to extend our results to higher-spin scenarios with transfer matrix fusion.

\section{The unbalanced RLL relations}
\label{app:RLL}
\label{subapp:RLLfactorization}

In this Appendix, we present the unbalanced RLL relations that are used in the main text to analyze the non-Abelian symmetry at roots of unity (the algebra of non-commuting transfer matrices). More specifically, we derive families of relations involving the (semi)cyclic/$\tau_2$ L--operator  $\mathcal{L}(u,v,s)$, defined as in eq. \eqref{eq:Tc} above.
The generic form of the unbalanced RLL equations was described in the main text; however, in the following, it will turn out to be more practical to rewrite them in the modified form,
\begin{equation}
    \check{\mathcal{R}}_{a,b} \mathcal{L}_a^{(1)} \mathcal{L}_b^{(2)} = \mathcal{L}_a^{(3)} \mathcal{L}_b^{(4)} \check{\mathcal{R}}_{a,b} \,,
\label{eq:nonabelianRcheckLL} 
\end{equation}
where we use $\check{\mathcal{R}}_{a,b}=\mathcal{P}_{a,b}\mathcal{R}_{a,b}$, with $\mathcal{P}_{a,b}:|i\rangle_a |j\rangle_b \to |j\rangle_a |i\rangle_b$ being the permutation operator that exchanges the two auxiliary spaces.
Eq.~\eqref{eq:nonabelianRcheckLL} is obtained from the original equation presented in the main text after a relabeling of the auxiliary spaces. 
In pictures, 
\be 
\newcommand\rad{0.35}
\begin{tikzpicture}[baseline=(current  bounding  box.center),scale=1.]
\draw[line width=1,red, rounded  corners=10pt] (0,-0.5) -- (0.5,-0.5) -- (1,0);
\draw[line width=1,-<,>=latex,green!40!black, rounded  corners=10pt] (1,0) -- (1.5,0.5) -- (2.5,0.5);
\draw[line width=1,blue, rounded  corners=10pt] (0,0.5) -- (0.5,0.5) -- (1,0);
\draw[line width=1,-<,>=latex,orange, rounded  corners=10pt] (1,0) -- (1.5,-0.5) -- (2.5,-0.5);
\draw[line width=1,>-,>=latex] (2,-1) -- (2,1);

\draw[gray,fill=gray!10,thick]  (1,0) circle (\rad); \node at (1,0) {$ \check{\cal R}$};
\draw[gray,fill=gray!10,thick]  (2,0.5) circle (\rad); \node at (2,0.5) {${\cal L}_a^{(1)}$};
\draw[gray,fill=gray!10,thick]  (2,-0.5) circle (\rad); \node at (2,-0.5) {${\cal L}_b^{(2)}$};
\end{tikzpicture}
\quad 
=
\quad
\begin{tikzpicture}[baseline=(current  bounding  box.center),scale=1.]
\draw[line width=1,red, rounded  corners=10pt] (0.5,-0.5) -- (1.5,-0.5) -- (2,0);
\draw[line width=1,-<,>=latex,green!40!black, rounded  corners=10pt] (2,0) -- (2.5,0.5) -- (3.,0.5);
\draw[line width=1,blue, rounded  corners=10pt] (0.5,0.5) -- (1.5,0.5) -- (2,0);
\draw[line width=1,-<,>=latex,orange, rounded  corners=10pt] (2,0) -- (2.5,-0.5) -- (3.,-0.5);
\draw[line width=1,>-,>=latex] (1,-1) -- (1,1);

\draw[gray,fill=gray!10,thick]  (2,0) circle (\rad); \node at (2,0) {$\check{\cal R}$};
\draw[gray,fill=gray!10,thick]  (1,0.5) circle (\rad); \node at (1,0.5) {${\cal L}_a^{(3)}$};
\draw[gray,fill=gray!10,thick]  (1,-0.5) circle (\rad); \node at (1,-0.5) {${\cal L}_b^{(4)}$};
\end{tikzpicture} 
\,.
\ee

\subsection{Elementary intertwiners}
\label{subapp:intertwiners}

Let us start by rewriting the cyclic/$\tau_2$ Lax operator in factorized form. Denoting $t={(u+v)}/{2}$ and $w = {(u-v)}/{2}$, 
it is straightforward to check that
\begin{equation}
\begin{split}
    \mathcal{L}_{\aux , j} (u,v,s) & = \begin{pmatrix}
        1 & {\color{gray} 0} \\ {\color{gray} 0} & X_a
    \end{pmatrix}_j \begin{pmatrix}
        e^{-t} & e^t \\ e^t & -e^{-t}
    \end{pmatrix}_j  \begin{pmatrix}
        1 & {\color{gray} 0} \\ {\color{gray} 0} & Z_a
    \end{pmatrix}_j \begin{pmatrix}
        e^s & {\color{gray} 0} \\ {\color{gray} 0} & e^{-s}
    \end{pmatrix}_j \begin{pmatrix}
        e^{-w} & e^w \\ e^w & -e^{-w}
    \end{pmatrix}_j  \begin{pmatrix}
        1 & {\color{gray} 0} \\ {\color{gray} 0} & \omega X_a^{-1}
    \end{pmatrix}_j 
    \\ 
    &
    := \big\{ t , s , w \big\}_{\aux, j} \; ,
\end{split}
\end{equation}
where we have introduced the short-hand notation $\{t,s,w\}_{\aux,j}$.
Our strategy for deriving intertwining relations between generic pairs of Lax operators follows the same line as Refs.~\cite{MLP_2021, Weston:2024dml}, namely, we will decompose the intertwining relations as sequences of elementary operations. Let us start by listing these elementary operations, which are also summarized in Fig.~\ref{fig:intertwiners}.
The first building block corresponds to a pair of operators acting in $a=\mathbb{C}^N$, 
\begin{equation}
    \mathsf{M}_a (x) \equiv \sum_{n=1}^N \prod_{i=1}^n \frac{e^{2x}  - \varepsilon^{-1} \omega^{i-1}}{ 1- \varepsilon^{-1} \omega^{i}   e^{2x} } Z_a^n \; ,
    \qquad
  \tilde{\mathsf{M}}_a (x)  \equiv \sum_{n=1}^N \prod_{i=1}^n \frac{e^{2x} \omega^{i-1} - \varepsilon^{-1} }{ \omega^{i} - \varepsilon^{-1} e^{2x} } Z_a^n \;.
\end{equation}
Here and in the following we use $\varepsilon = e^{\frac{\rmi \pi}{N}}$.
These operators are unnormalized inverses of each other (namely $\mathsf{M}_a(x) \tilde{\mathsf{M} }_a(x)=\tilde{\mathsf{M}}_a(x) \mathsf{M}_a(x) \propto \mathrm{id}_a$), and intertwine $\{t,0,w\}$ and $\{w,0,t\}$ as follows
\begin{align}
  & \mathsf{M}_a(t-w) \big\{t,0,w\big\}_{\aux,j} =  \varkappa_j \cdot \big\{w,0,t\big\}_{\aux,j} \cdot \varkappa^{-1}_j  \mathsf{M}_a(t-w) \; ,
 \nonumber 
 \\
 &\tilde{\mathsf{M}}_a(t-w) \varkappa_j \cdot \big\{ w,0,t \big\}_{\aux,j} \cdot \varkappa_j^{-1}  = \big\{ t,0,w \big\}_{\aux,j}  \tilde{\mathsf{M}}_a(t-w)\; , 
 \qquad 
 \varkappa_j \equiv \begin{pmatrix}
        1 & {\color{gray} 0} \\ {\color{gray} 0} & \varepsilon
    \end{pmatrix}_j
    \; .
\label{eq:intertwinerM}
\end{align}
Relations \eqref{eq:intertwinerM} can be directly checked by writing all sides as $2\times 2$ matrices, and using the algebra of $X_a, Z_a$.

The other elementary intertwiners that we define act on the tensor product $a\otimes b$ of two auxiliary spaces,  
\begin{eqnarray}
    \mathsf{S}_{a,b}(t) &\equiv& \sum_{n=1}^N \prod_{i=1}^n \frac{\varepsilon^{-1}(e^{2t} - \omega^{i-1} )}{1-e^{2t} \omega^i } (\omega X_a^{-1} X_b)^n \; ,
    \\
     \mathsf{T}_{a,b}(w) &\equiv& \sum_{n=1}^N \prod_{i=1}^n \frac{\varepsilon(e^{2w} - \omega^{i-1} )}{1-e^{2w}\omega^i }  (\omega X_a^{-1} X_b)^n  \; ,
    \\
    \mathsf{K}_{a,b} (s) &\equiv& \sum_{n=0}^{N-1} \prod_{i=1}^{n}\frac{e^{2s} - \omega^{i-1} }{e^{2s} - \omega^i} (\omega X_a^{-1} X_b)^n \; .
\label{eq:intertwinerKdef}
\end{eqnarray}
These operators have (unnormalized) inverses obtained by negating their argument, $\mathsf{K}_{a,b}(s)\mathsf{K}_{a,b}(-s)\propto \mathrm{id}_{a,b}$ (and similarly after replacing $\mathsf{S}$ by  $\mathsf{T}$ or $\mathsf{K}$), and can be checked to obey the following relations 
\begin{align}
   & \mathsf{S}_{a,b}(t-t')  ~\varkappa_j \cdot \big\{w,0,t\big\}_{\aux,j} \cdot \varkappa_j^{-1}   
    \big\{t',s,w'\big\}_{b,j}
      =  \varkappa_j \cdot \big\{w,0,t'\big\}_{a,j}
    \cdot \varkappa_j^{-1}  \big\{t,s,w'\big\}_{b,j}
    ~
     \mathsf{S}_{a,b}(t-t') \; .
     \label{eq:intertwinerS}
\\
 &\mathsf{T}_{a,b}(w-w')
 \big\{ t,s,w \big\}_{a,j}
 \varkappa_j \cdot 
  \big\{ w',0,t' \big\}_{b,j}
  \cdot \varkappa_j^{-1} 
  = 
    \big\{ t,s,w' \big\}_{a,j}
   \varkappa_j \cdot 
    \big\{ w,0,t' \big\}_{b,j}
   \cdot \varkappa_j^{-1}  
   ~ \mathsf{T}_{a,b}(w-w') \; .
   \label{eq:intertwinerT}
\\
 & \mathsf{K}_{a,b}(s)~\big\{t,s,0\}_a \big\{ 0,s',w \big\}_b 
=  
    \big\{t,0,0\}_a \big\{ 0,s+s',w \big\}_b 
  \mathsf{K}_{a,b} (s) \; .
   \label{eq:intertwinerK}
\end{align}
\begin{figure}\centering
\begin{tikzpicture}
\node at (0,0) {$\big\{ {\color{blue}t},0,{\color{blue}w} \big\}$};
    \draw[<->, blue]
(-0.35,0.3) to[out=60,in=120] (0.35,0.3);
\node [blue] at (0,0.8) {$\mathsf{M}$};

\begin{scope}[xshift=4cm]
\node at (0,0) {$\big\{ w,0,{\color{blue}t} \big\} \big\{{\color{blue}t'},s , w' \big\} $};
    \draw[<->, blue]
(-0.4,0.3) to[out=60,in=120] (0.2,0.3);
\node [blue] at (-0.1,0.8) {$\mathsf{S}$};
\end{scope}

\begin{scope}[xshift=8cm]
\node at (0,0) {$\big\{ t, s, {\color{blue}w} \big\} \big\{ {\color{blue}w'},0,t' \big\} $};
    \draw[<->, blue]
(-0.35,0.3) to[out=60,in=120] (0.25,0.3);
\node [blue] at (-0.05,0.8) {$\mathsf{T}$};
\end{scope}

\begin{scope}[xshift=12cm]
\node at (0,0) {$\big\{ t, {\color{blue}s},0 \big\} \big\{ 0,s',w \big\} $};
    \draw[double,->, blue]
(-0.6,0.3) to[out=60,in=120] (0.45,0.3);
\node[blue] at (0.,0.8) {$\mathsf{K}$};
\end{scope}

\end{tikzpicture}
\caption{
Summary of the basic intertwining relations \eqref{eq:intertwinerM}, \eqref{eq:intertwinerS}, \eqref{eq:intertwinerT}, \eqref{eq:intertwinerK}.
}
\label{fig:intertwiners}
\end{figure}

\subsection{The unbalanced RLL relations}
\label{subapp:nonabelianRLL}

By constructing sequences of the five intertwiners $\mathsf{M}, \,\tilde{\mathsf{M}}, \, \mathsf{S}, \, \mathsf{T}, \, \mathsf{K}$, we can now readily derive the various unbalanced RLL relations obeyed by the cyclic/$\tau_2$ Lax operators. 
The first such relation is given by 
\begin{equation}
\begin{split}
&
    \check{\mathcal{R}}_{ab} ~ \mathcal{L}_{aj} (u,v,s) \mathcal{L}_{bj} (0,0,0) = \mathcal{L}_{aj} \left( \tfrac{u+v}{2} , \tfrac{u+v}{2} , s \right) \mathcal{L}_{bj} \left( \tfrac{u-v}{2} , \tfrac{v-u}{2} , 0 \right) \check{\mathcal{R}}_{ab} \; ,
\\
&
    \check{\mathcal{R}}_{ab} \equiv \tilde{\mathsf{M}}_b (\tfrac{u-v}{2}) \mathsf{T}_{a,b} (\tfrac{u-v}{2}) \mathsf{M}_b( 0  ) \; ,
    \end{split}
\end{equation}
%
and follows from the sequence,
\begin{equation}
\begin{split}
 \mathcal{L}_{aj} (u,v,s) \mathcal{L}_{bj} (0,0,0) &=
     \{ t,s,w\}_{a,j} \cdot \{ 0,0,0\}_{b,j}
     \\&
     \xrightarrow{\mathsf{M}_b (0 )} \{ t,s,w\}_{a,j} \cdot \varkappa_j \{ 0,0,0\}_{b,j} \varkappa^{-1}_j 
     \\ 
     &\xrightarrow{\mathsf{T}_{a,b} (\tfrac{u-v}{2})}  \{ t,s,0\}_{a,j} \cdot \varkappa_j \{ w,0,0\}_{b,j} \varkappa^{-1}_j 
     \\ 
     &\xrightarrow{\tilde{\mathsf{M}}_b (\tfrac{u-v}{2})}  \{ t,s,0\}_{a,j} \cdot \{ 0,0,w\}_{b,j}  
     = \mathcal{L}_{aj} \left( \tfrac{u+v}{2} , \tfrac{u+v}{2} , s \right) \mathcal{L}_{bj} \left( \tfrac{u-v}{2} , \tfrac{v-u}{2} , 0 \right)
     \; ,
    \label{eq:sequenceSB11}
\end{split}
\end{equation}
with $t = \tfrac{u+v}{2}$ and $w=\tfrac{u-v}{2}$. 
By using the train argument as described in the main text, this yields the first step of the factorization for the cyclic/$\tau_2$ transfer matrix
\begin{equation}
    \bm T_{\rm c}(u,v,s) \bm T_{\rm c}(0,0,0)  = \bm T_{\rm c} \left(\tfrac{u+v}{2},\tfrac{u+v}{2},s \right) \bm T_{\rm c} \left( \tfrac{u-v}{2}, \tfrac{v-u}{2},0 \right) \; .
    \label{eq:TcuvsTc0001} 
\end{equation}

A similar relation is
\begin{equation}
\begin{split}
 &   \check{\mathcal{R}}_{ab} \mathcal{L}_{aj} \left( \tfrac{u+v}{2} , \tfrac{u+v}{2} , s \right) \mathcal{L}_{bj} \left( \tfrac{u-v}{2} , \tfrac{v-u}{2} , 0 \right)  = \mathcal{L}_{aj} (0,0,0) \mathcal{L}_{bj} (u,v,s) ~ \check{\mathcal{R}}_{ab}  \; ,
\\
 &   \check{\mathcal{R}}_{ab} \equiv \tilde{\mathsf{M}}_a (0) \mathsf{S}_{a,b} ({\tfrac{u+v}{2}}) \mathsf{M}_a (\tfrac{u+v}{2}) \mathsf{K}_{a,b} (s) \; .
\end{split}
\end{equation}
The derivation of the intertwiner $\check{\mathcal{R}}_{ab}$ is analogous to the sequence in \eqref{eq:sequenceSB11}. 
By using the train argument and combining with \eqref{eq:TcuvsTc0001}, we have
\begin{equation}
    \bm T_{\rm c}(u,v,s) \bm T_{\rm c}(0,0,0)  = \bm T_{\rm c} \left(\tfrac{u+v}{2},\tfrac{u+v}{2},s \right) \bm T_{\rm c} \left( \tfrac{u-v}{2}, \tfrac{v-u}{2},0 \right) = \bm T_{\rm c}(0,0,0) \bm T_{\rm c}(u,v,s)\; ,
    \label{eq:TcuvsTc000commute}
\end{equation}
i.e. $\bm T_{\rm c}(0,0,0)$ commutes with all $\tau_2$ transfer matrices with arbitrary parameters.

Another relation is 
\begin{equation}
\begin{split}
&    \check{\mathcal{R}}_{ab} \mathcal{L}_{aj} \left( \tfrac{u+v}{2} , \tfrac{u+v}{2} , 0 \right) \mathcal{L}_{bj} \left( \tfrac{u'+v'}{2} , \tfrac{u'+v'}{2} , 0 \right) = \mathcal{L}_{aj} \left( \tfrac{u'+v'}{2} , \tfrac{u'+v'}{2} , 0 \right) \mathcal{L}_{bj} \left( \tfrac{u+v}{2} , \tfrac{u+v}{2} , 0 \right) \check{\mathcal{R}}_{ab} \; ,
\\
&    \check{\mathcal{R}}_{ab} \equiv \tilde{\mathsf{M}}_a (\tfrac{u'+v'}{2} ) \mathsf{S}_{a,b} (\tfrac{u+v-u'-v'}{2}) \mathsf{M}_a (\tfrac{u+v}{2}  ) \; ,
\end{split}
\end{equation}
which implies that
\begin{equation}
    \left[ \bm T_+(u) , \bm T_+(v) \right] = 0 \; .
\label{eq:comTpTp}
\end{equation}

Similarly, we have
\begin{equation}
\begin{split}
&    \check{\mathcal{R}}_{ab} \mathcal{L}_{aj} \left( \tfrac{u-v}{2} , \tfrac{v-u}{2} , 0 \right) \mathcal{L}_{bj} \left( \tfrac{u'-v'}{2} , \tfrac{v'-u'}{2} , 0 \right) = \mathcal{L}_{aj} \left( \tfrac{u'-v'}{2} , \tfrac{v'-u'}{2} , 0 \right) \mathcal{L}_{bj} \left( \tfrac{u-v}{2} , \tfrac{v-u}{2} , 0 \right) \check{\mathcal{R}}_{ab} \; ,
\\
&
   \check{\mathcal{R}}_{ab} \equiv \tilde{\mathsf{M}}_b (\tfrac{v-u}{2} ) \mathsf{T}_{a,b} ( \tfrac{u-v-u'+v}{2} ) \mathsf{M}_b (\tfrac{v'-u'}{2}  ) \;,
\end{split}
\end{equation}
which implies that
\begin{equation}
    \left[ \bm T_-(u) , \bm T_-(v) \right] = 0 \; .
    \label{eq:comTmTm}
\end{equation}

Finally,
\begin{equation}
\begin{split}
&    \check{\mathcal{R}}_{ab} \mathcal{L}_{aj} \left( \tfrac{u+v}{2} , \tfrac{u+v}{2} , 0 \right) \mathcal{L}_{bj} \left( \tfrac{u'-v'}{2} , \tfrac{v'-u'}{2} , 0 \right) = \mathcal{L}_{aj} \left( \tfrac{u'-v'}{2} , \tfrac{v'-u'}{2} , 0 \right) \mathcal{L}_{bj} \left( \tfrac{u+v}{2} , \tfrac{u+v}{2} , 0 \right) \check{\mathcal{R}}_{ab} \; ,
\\
&
    \check{\mathcal{R}}_{ab} \equiv \tilde{\mathsf{M}}_a (\tfrac{v'-u'}{2}) \mathsf{S}_{a,b} (\tfrac{u+v}{2} ) \mathsf{M}_a(\tfrac{u+v+v'-u'}{2}) \tilde{\mathsf{M}}_b(0) \mathsf{T}_{a,b} (\tfrac{v'-u'}{2}) \mathsf{M}_b \big(\tfrac{v'-u'}{2} \big) \; ,
\end{split}
\end{equation}
which implies that
\begin{equation}
    \left[ \bm T_+(u) , \bm T_-(v) \right] = 0 \; .
    \label{eq:comTpTm}
\end{equation}

\section{Relating the transfer matrices with the Onsager algebra}
\label{app:Transfer_Onsager}

\subsection{Basics about the Onsager algebra}
\label{subapp:Onsager_algebra}

The Onsager algebra~\cite{Onsager, El-Chaar_2012} is defined from two infinite families of generators, $\mathbf{A}_m$, $\mathbf{G}_l$ ($m,l\in \mathbb{Z}$), with $\mathbf{G}_{-l}=-\mathbf{G}_l$, obeying the following commutation relations 
\be
[\mathbf{A}_l, \mathbf{A}_m]=4\mathbf{G}_{l-m}  \,,
\qquad
[\mathbf{G}_l, \mathbf{A}_m]=2\mathbf{A}_{l+m}-2\mathbf{A}_{m-l}  \,,
\qquad 
[\mathbf{G}_l, \mathbf{G}_m]=0  \,,
\label{eq:Onsagercomrel} \,.
\ee 
There are many other possible ways to organize the generators. 
Onsager's original presentation, displayed above, gives a ``symmetric'' role to $\Ao$ and $\Ai$, related by Kramers-Wannier duality (while the $\mathbf{G}_l$s are odd under duality).
We here briefly review an alternative presentation, introduced in Ref.~\cite{Vernier:2018han}, which in contrast organizes the generators according to their behavior with respect to the $U(1)$ charge $\Ao$. The generators are now $\mathbf{Q}^\alpha_m$, $\alpha\in\{0,\pm\}$, $m\in \mathbb{N}$, linearly related to the former through
\be \mathbf{Q}^0_m=\frac{\mathbf{A}_m+\mathbf{A}_{-m}}{2}\,, \qquad 
\mathbf{Q}^\pm_m = \frac{ \mathbf{A}_m-\mathbf{A}_{-m} \pm 2\mathbf{G}_m}{4} \,, 
\ee
(and conversely, $\mathbf{A}_m=\mathbf{Q}_m^0 + \mathbf{Q}_m^+ + \mathbf{Q}_m^-$ and $\mathbf{G}_m=\mathbf{Q}_m^+ - \mathbf{Q}_m^-$, see Ref.~\cite{Miao_2021} for a proof of the isomorphism). Each of the three families $\{\mathbf{Q}_m^\alpha\}$ for $\alpha=0,+,-$ forms an abelian subalgebra, each with a different behaviour with respect to $\Ao$:  $\mathbf{Q}_m^{0}$ commutes with it, while $\mathbf{Q}_m^\pm$ raise (lower) it by $\pm 4$.

\subsection{Proof of the first Dolan--Grady relation}
\label{subapp:proofDG1}

We start by introducing in the auxiliary space $a=\mathbb{C}^N$ raising and lowering nilpotent operators $S_a^\pm$ such that $X_a^{\pm 1} = S_a^\pm + (S_a^\mp)^{N-1}$, namely which in the basis \eqref{eq:Zbasis} act as $S_a^\pm |n\rangle = |n\pm 1\rangle$, see the discussion above eq.~\eqref{eq:Tn}. Defining further $U_a |n\rangle  = e^{\alpha n} |n\rangle$ , we thus have $U_a S_a^\pm U_a^{-1} = e^{\pm \alpha} S_a^\pm$. This allows us to write 
\begin{align} 
&(e^{\frac{\alpha}{2} \sigma_j^z}  U_a) \cdot  \mathcal{L}_{a,j}(u,v,s) \cdot
(e^{-\frac{\alpha}{2} \sigma_j^z}    U_a^{-1})
 \nonumber 
 \\
& =  \left(
\begin{array}{cc}
e^{s-u} + e^{u-s} Z_a &  \omega (e^{s-v} - e^{v-s} Z_a)(S_a^- + e^{N\alpha}(S_a^+)^{N-1}) \\
 (S_a^+ + e^{-N\alpha}(S_a^-)^{N-1})(e^{v+s} -  e^{-v-s} Z_a) & \omega e^{u+s} + e^{-u-s} Z_a
\end{array} 
\right)_j 
  \,.
  \label{eq:U1gauge}
\end{align} 
Noting that for $v=s=0$ the $(S_a^\pm)^{N-1}$ terms vanish from the off-diagonal terms, as a result of $(1-Z_a)(S_a^+)^{N-1}=(S_a^-)^{N-1}(1-Z_a)=0$, we see that the r.h.s. of \eqref{eq:U1gauge} recovers the Lax operator $\mathcal{L}_{a,j}(u,0,0)$. 
Using this identity repeatedly yields
\be 
e^{\alpha S^z} \bm T_{\rm c}(u,0,0) e^{-\alpha S^z} = \mathrm{Tr}_{a}(U_a^{-1} \mathcal{L}_{a,1}(u,0,0)U_a \ldots U_a^{-1} \mathcal{L}_{a,L}(u,0,0)U_a) = \bm T_{\rm c}(u,0,0) \,,
\ee
proving commutation of the matrix $T(u)=\bm T_{\rm c}(u,0,0)$ with the global magnetization $S^z$. 

Taking into account now the derivative of the cyclic transfer matrix with respect to $s$ at $s=0$, we obtain from the same identity
\be 
e^{\alpha S^z} \partial_s \bm T_{\rm c}(u,0,s)|_{s=0} e^{-\alpha S^z}  
= \sum_{j=1}^L \mathrm{Tr}_{a}(\mathcal{L}_{a,1}(u,0,0)\ldots  \widetilde{(\partial_s\mathcal{L})}_{a,j}(u,\alpha) \ldots  \mathcal{L}_{a,L}(u,0,0))
\ee 
where 
\begin{equation} 
\begin{split}
 \widetilde{(\partial_s\mathcal{L})}_{a,j}(u,\alpha) 
 &= 
 (e^{\frac{\alpha}{2} \sigma_j^z} ~ U_a)   \cdot \partial_s\mathcal{L}(u,0,s)_{a,j}|_{s=0} \cdot
(e^{-\frac{\alpha}{2} \sigma_j^z}  ~  U_a^{-1})
\\ 
&= \partial_s \mathcal{L}_{a,j}(u) + 2\cosh(N\alpha)  \left(
\begin{array}{cc}
0 &  \omega (S_a^+)^{N-1} \\
 (S_a^-)^{N-1} &  0
\end{array} 
\right)_j 
+ 
2\sinh(N\alpha)
\left(
\begin{array}{cc}
0 &  \omega (S_a^+)^{N-1} \\
 -(S_a^-)^{N-1}&  0
\end{array} 
\right)_j  \,.
\end{split}
\label{eq:toDG1}
\end{equation}
At the expense of yet another rotation in the auxiliary space, we similarly show that
\be 
\partial_v \bm T_{\rm c}(u,0,v)|_{v=0}   
= \sum_{j=1}^L \mathrm{Tr}_{a}(\mathcal{L}_{a,1}(u,0,0)\ldots  \left(
\begin{array}{cc}
0 &  -2\omega (S_a^+)^{N-1} \\
 2(S_a^-)^{N-1}&  0
\end{array} 
\right)_j 
 \ldots  \mathcal{L}_{a,L}(u,0,0)) 
\label{eq:DvT}
\,.
\ee 
Expanding the l.h.s. of \eqref{eq:toDG1} with the help of the Baker--Campbell--Hausdorff formula (Hadamard's conjugation formula), we recover at $u=0$ the first Dolan--Grady equation.
Furthermore, we get from comparing with \eqref{eq:DvT} the useful relation 
\be 
\partial_v \bm T_{\rm c}(u,0,v)|_{v=0}   
= \frac{-1}{4}\left[\Ao ,~ \partial_s \bm T_{\rm c}(u,0,s)|_{s=0}   
\right]
\label{eq:usefulidentitybis}
\ee  
which we will use below to relate $\bm T_{\pm}(u)$ to the Onsager generators, see Sec.~\ref{subapp:Tpm_expansion}.

\begin{figure}
\newcommand\rad{0.35}

\centering
\begin{tikzpicture}[baseline=(current  bounding  box.center),scale=1.2]
\draw[line width=1,green!40!black, rounded  corners=10pt,-<,>=latex] (-0.5,0.75) -- (1.5,0.75) -- (6,0.75);
\draw[line width=1,orange, rounded  corners=10pt,-<,>=latex] (-0.5,-0.75) -- (1.5,-0.75) -- (6,-0.75);
\foreach \x in {0.25,1.5,...,6} 
{
\draw[line width=1,>-,>=latex] (\x,-1.35) -- (\x,1.2);
\draw[gray,fill=gray!10,thick]  (\x,0.75) circle [x radius= \rad, y radius= \rad]; \node at (\x,0.75) {\footnotesize  ${\cal L}_a(s)$};
\draw[gray,fill=gray!10,thick]  (\x,-0.75) circle [x radius= \rad, y radius= \rad]; \node at (\x,-0.75) {\footnotesize ${\cal L}_b(-s)$};
}
\draw[white,fill=white,thick]  (2.75,0) circle (0.5*\rad); \node at (2.75,0) { $\sigma^z$};
\end{tikzpicture}
\quad
=
\quad
\begin{tikzpicture}[baseline=(current  bounding  box.center),scale=1.3]
\draw[line width=1,blue, rounded  corners=10pt] (-0.5,0.75) -- (1.75,0.75) -- (2,0);
\draw[line width=1,red, rounded  corners=10pt] (-0.5,-0.75) -- (1.75,-0.75) -- (2,0);
\draw[line width=1,red, rounded  corners=10pt,>-,>=latex] (6,-0.75) -- (3.75,-0.75) -- (3.5,0);
\draw[line width=1,blue, rounded  corners=10pt,>-,>=latex] (6,0.75) -- (3.75,0.75) -- (3.5,0);
\draw[line width=1,green!40!black, rounded  corners=10pt] (2,0) -- (2.25,0.75) -- (3.25,0.75) -- (3.5,0);
\draw[line width=1,orange, rounded  corners=10pt] (2,0) -- (2.25,-0.75) -- (3.25,-0.75) -- (3.5,0);
\draw[gray,fill=gray!10,thick] (2.,0) circle (\rad); \node at (2,0) {\footnotesize $\check{\mathcal{R}}_{a,b}$};
\draw[gray,fill=gray!10,thick] (3.5,0) circle (\rad); \node at (3.5,0) {\footnotesize $\check{\mathcal{R}}_{a,b}^{-1}$};

\draw[line width=1,>-,>=latex] (2.75,-1.35) -- (2.75,1.35);
\draw[gray,fill=gray!10,thick]  (2.75,0.75) circle [x radius= \rad, y radius= \rad]; \node at (2.75,0.75) {\footnotesize  ${\cal L}_a(s)$};
\draw[gray,fill=gray!10,thick]  (2.75,-0.75) circle [x radius= \rad, y radius= \rad]; \node at (2.75,-0.75) {\footnotesize  ${\cal L}_b(-s)$};

\foreach \x in {0.25,1.25,4,5.25} 
{
\draw[line width=1,>-,>=latex] (\x,-1.35) -- (\x,1.35);
\draw[gray,fill=gray!10,thick]  (\x,0.75) circle [x radius= \rad, y radius= \rad]; \node at (\x,0.75) {\footnotesize  ${\cal L}_a(0)$};
\draw[gray,fill=gray!10,thick]  (\x,-0.75) circle [x radius= \rad, y radius= \rad]; \node at (\x,-0.75) {\footnotesize  ${\cal L}_b(0)$};
}
\draw[white,fill=white,thick]  (2.75,0) circle (0.5*\rad); \node at (2.75,0) { $\sigma^z$};
\end{tikzpicture}

\caption{
Pictorial representation of the calculation made in Sec. \ref{subapp:proofDG2}.
}
\label{fig:doublerow}
\end{figure}
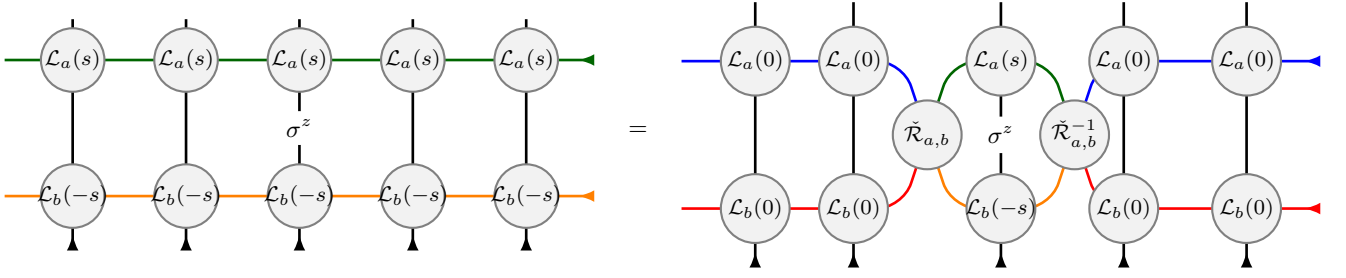

\subsection{Proof of the second Dolan--Grady relation}
\label{subapp:proofDG2}

We now turn to establishing the second Dolan--Grady relation. 
Consider the product 
\be 
\bm T_{\rm c}(0,0,s) ~\Ao~ \bm T_{\rm c}(0,0,-s) 
=
\frac{2}{N} \sum_{j=1}^L \mathrm{Tr}_{a \otimes b} \left(
\mathcal{L}_{a,b,1}(s)
\ldots
\hat{\mathcal{L}}_{a,b,j}(s)
\ldots 
\mathcal{L}_{a,b,L}(s)\right)
 \,, 
\ee 
where the two-row matrices $\mathcal{L}_{a,b,j}(s)$ and $\hat{\mathcal{L}}_{a,b,j}(s)$ are defined as 
\be 
\mathcal{L}_{a,b,j}(s) = \mathcal{L}_{a,j}(0,0,s)\mathcal{L}_{b,j}(0,0,-s) 
\,,
\qquad
\hat{\mathcal{L}}_{a,b,j}(s) = \mathcal{L}_{a,j}(0,0,s) \sigma_j^z \mathcal{L}_{b,j}(0,0,-s)  \,,
\ee 
matrix products being taken in the physical space $j$ while the entries of $\mathcal{L}_{a,j}(0,0,s)$ (resp. $\mathcal{L}_{b,j}(0,0,s)$) act respectively on each tensorand of the doubled auxiliary space. See Fig.~\ref{fig:doublerow} for a pictorial representation.
Using the RLL equation \eqref{eq:intertwinerK} with $t=w=0$, $s'=-s$ and repeating the train argument allows us to rewrite 
\be 
\bm T_{\rm c}(0,0,s) ~\Ao~ \bm T_{\rm c}(0,0,-s) 
=
\frac{2}{N} \sum_{j=1}^L \mathrm{Tr}_{a \otimes b} \left(
\mathcal{L}_{a,b,1}(0)
\ldots
\widetilde{\mathcal{L}}_{a,b,j}(s)
\ldots 
\mathcal{L}_{a,b,L}(0)\right)
 \,, 
 \label{eq:TsAoTs}
\ee 
where
\be 
\widetilde{\mathcal{L}}_{a,b,j}(s) = \check{\mathcal{R}}_{a,b}(-s) 
\mathcal{L}_a(0,0,s) \sigma_j^z \mathcal{L}_b(0,0,-s)
\check{\mathcal{R}}_{a,b}(s)\,,
\ee 
with $\check{\mathcal{R}}_{a,b}$ proportional to $\mathsf{K}_{a,b}(s)$ defined in \eqref{eq:intertwinerKdef}, obeying $\check{\mathcal{R}}_{a,b}(s)\check{\mathcal{R}}_{a,b}(-s) = 1$.
In the matrix product, the only operators that do not commute with $\check{\mathcal{R}}_{a,b}(s)$ are $Z_a$ and $Z_b$, and they appear through the combinations $e^{2s} Z_a$, $e^{-2s} Z_b$, $Z_a Z_b$. Moreover, the only dependence in $s$ is through these combinations.
Using the identities : 
\be 
\begin{split}
\check{\mathcal{R}}_{a,b}(s) (e^{2s} Z_a) \check{\mathcal{R}}_{a,b}(-s) &= (1- (e^{2N s}-1) \Pi_{a,b}) Z_a \\
\check{\mathcal{R}}_{a,b}(s) (e^{-2s} Z_b) \check{\mathcal{R}}_{a,b}(-s) &= Z_b (1- (e^{-2N s}-1) \Pi_{a,b})  \\
\check{\mathcal{R}}_{a,b}(s) (Z_a Z_b) \check{\mathcal{R}}_{a,b}(-s) &= Z_a Z_b  \,,
\end{split}
\ee 
where we have introduced the projector $\Pi_{a,b}=\frac{1}{N}\sum_{k=0}^{N-1} X_a^k X_b^{-k}$, we see that the only dependence in $s$ of the matrix product \eqref{eq:TsAoTs} is as a linear combination of $e^{2N s}$, $e^{- 2N s}$, and an $s$-independent term. 
Introducing the exponential form of $\bm T_{\rm c}(0,0,s)=T(0) e^{\frac{s N}{2}\Ai}$ and comparing with the Baker--Campbell--Hausdorff formula, we then recover the second Dolan--Grady relation. 

\subsection{Expansion of \texorpdfstring{$\bm T_\pm(u)$}{Tpm(u)} in terms of the Onsager generators}
\label{subapp:Tpm_expansion}

The factorization formula for the cyclic/$\tau_2$ transfer matrix, given in eq.~\eqref{eq:Ttau2factorization} of the main text, fixes that 
\be 
\bm T_+(u) \bm T_-(u) 
= T(0)\bm T_{\rm c}(u,u,0) 
= T(0)^2 T(u)
= T(0)^2 e^{\mathrm{i} \mathcal{H}(u)} \,, 
\ee 
therefore we can freely define $\bm T_\pm(u) =  T(0) e^{\pm\frac{ 1}{2}\mathcal{G}(u)+ \frac{i}{2} \mathcal{H}(u)}$, as in the main text. Furthermore $[\mathcal{G}(u),\mathcal{G}(v)]=[\mathcal{H}(u),\mathcal{H}(v)]=[\mathcal{G}(u),\mathcal{H}(v)]$ for all $u,v$, as a result of the commutation of the families $\bm T_\pm(u)$, Eqs.~\eqref{eq:comTpTp}, \eqref{eq:comTmTm}, \eqref{eq:comTpTm}.
Plugging this expression (as well as the exponential form of $\bm T_{\rm c}(u,0,s)$ into the identity \eqref{eq:usefulidentitybis} results in
\be  
    \mathcal{G}'(u) = -\frac{N}{4} [\Ao ,  e^{\frac{1}{2}\mathcal{G}(u)} \Ai e^{-\frac{1}{2}\mathcal{G}(u)}  ]  \,,
 \label{eq:identityG}
 \ee 
the identity given in main text.
Expanding $\mathcal{G}(u)$ in series with respect to $u$ (with $\mathcal{G}(0)=0$) shows that the first order is proportional to $[\Ai,\Ao]=4 \mathbf{G}_1$. More generally, at a given order $n$, the coefficient of in the l.h.s. involves commutators of the form $[\Ao,[\ldots , \Ai]]$, where $\ldots$ denotes (possibly iterated) commutations with the lower-order terms in $\mathcal{G}$. Since commuting any $\mathbf{G}_l$ with some $\mathbf{A}_m$ yields a linear combination of $\mathbf{A}_{n}$'s, which,  commuted again with $\Ao$, yield some linear combinations of $\mathbf{G}_l$'s, all orders in the series expansion of $\mathcal{G}(u)$ can be written as linear combinations of the $\mathbf{G}_l$'s. We are therefore left with the task of determining the coefficients of this series expansion.

An elegant way to check that eq.~\eqref{eq:Gseries} indeed furnishes the correct series expansion is to recast the Onsager commutation relations between $\mathbf{G}_l$ and $\mathbf{A}_m$ generators in terms of a linear operator $\Xi$ acting on the space of Onsager generators,
\be 
[\mathbf{G}_l,\mathbf{A}_m] = 2 \mathbf{A}_{m+l}-2 \mathbf{A}_{m-l}= 2 (\Xi^l-\Xi^{-l})\mathbf{A}_m \,, 
\qquad\qquad
\Xi^{\pm1} \mathbf{A}_m \equiv \mathbf{A}_{m\pm 1} \,.
\ee 
With the ansatz $\mathcal{G}(u) = \sum_{m\geq 1} \frac{\tanh(N u)^m}{m} \mathbf{G}_m$, we then have 
\bea 
 e^{\frac{1}{2}\mathcal{G}(u)} \Ai e^{-\frac{1}{2}\mathcal{G}(u)} 
  &=& \exp[\sum_{m\geq1}  \frac{\tanh(N u)^m}{m} (\Xi^m - \Xi^{-m})]\Ai \\
  &=& \frac{1-\tanh(N u)\Xi^{-1}}{1-\tanh(N u)\Xi} \Ai \\
  &=& \frac{1}{\cosh^2(N u)} \sum_{m\geq 1} \tanh(N u)^{m-1} \mathbf{A}_m - \tanh(N u) \Ao
\eea 
Commuting again with $\Ao$, 
\be 
[\Ao,  e^{\frac{1}{2}\mathcal{G}(u)} \Ai e^{-\frac{1}{2}\mathcal{G}(u)} ]
= \frac{-4}{\cosh^2(N u)} \sum_{m\geq 1} \tanh(N u)^{m-1} \mathbf{G}_m  = - \frac{4}{N}\mathcal{G'}(u) \,, 
\ee
indeed satisfying the condition \eqref{eq:identityG}. 
We therefore have proved the expression of $\mathcal{G}(u)$ as a series expansion in the generators $\mathbf{G}_m$.

\subsection{The nilpotent representations}
\label{subapp:nilpotent}

The transfer matrices built from nilpotent (``complex spin'') representations were defined above, eq.~\eqref{eq:Tn}.
Clearly, they reduce to $T(u)$ for $w=0$, but more generally they commute with the Hamiltonian, and with the (half)-integers spin transfer matrices. 
We further check 
\bea 
[\bm T_{\rm n}(u,w),\bm T_{\rm n}(u',w')]&=&0 
\label{eq:commutationnilpotent}
\\ 
\bm T_{\rm n}(u,w)T(0) &=&  
\bm T_{\rm n}\left(\frac{u+w}{2},\frac{u+w}{2}\right) 
\bm T_{\rm n}\left(\frac{u-w}{2},\frac{w-u}{2}\right) 
\equiv \bm T_{\rm L}(\frac{u+w}{2})\bm T_{\rm R}(\frac{u-w}{2}) 
\label{eq:factorizationnilpotent}
\eea  
As for the cyclic transfer matrices, the factorization \eqref{eq:factorizationnilpotent} follows from an RLL equation of the form
\begin{align} 
& 
\check{\mathcal{R}}_{a,b}(u,w)\mathcal{L}_a(u,w)\mathcal{L}_b(0,0) 
=
\mathcal{L}_a\left(\frac{u+w}{2},\frac{u+w}{2}\right)
\mathcal{L}_b\left(\frac{u-w}{2},\frac{w-u}{2}\right) 
\check{\mathcal{R}}_{a,b}(u,w)
\\
&\check{\mathcal{R}}_{a,b}(u,w)
=
\Omega_b(u,w)
~ 
\sum_{k=0}^{N-1}\left(
\prod_{j=1}^k \frac{e^{w-u}-e^{u-w}\omega^{1-j}}{1-\omega^{-j}} \right)(S^-_a  S^+_b)^k \,,
\label{eq:RLLnilpotent}
\end{align} 
where the matrix $\Omega_b$ acts diagonally in the $Z$-basis \eqref{eq:Zbasis} of $b=\mathbb{C}^N$ as 
$ \Omega_b(u,w)|n\rangle=\left(
\prod_{i=1}^{n-1} \frac{1-\omega^{-i}}{e^{w-u}-e^{u-w}\omega^{-i}} \right) |n\rangle$.
Using this factorization, it is then easy to work out the relation with the Onsager generators: with the same reasoning as for the cyclic representations, we can freely define 
\be 
\bm T_{\mathrm{L}/\mathrm{R}}(u)=
T(0) \exp\left({\pm \frac{1}{2}\mathcal{Q}(u) + \frac{i}{2} \mathcal{H}(u) } \right) \,,
\ee 
where $\mathcal{H}(u)$ is the central generating function appearing in $T(u)=\bm T_{\rm}(u,0)$, while $\mathcal{Q}(u)$ is yet to be computed. We now turn to the derivative $\partial_w \bm T_{\rm n}(u,w)|_{w=0}$. On the one hand, using the factorized form \eqref{eq:factorizationnilpotent},
\be 
\partial_w \bm T_{\rm n}(u,w)|_{w=0}  = \frac{1}{2} \mathcal{Q}'\left(\tfrac{u}{2}\right) T(u) \,.
\ee
On the other hand, in matrix product form,
\be 
\partial_w \bm T_{\rm n}(u,w)|_{w=0}   
= \sum_{j=1}^L \mathrm{Tr}_{a}(\mathcal{L}_{a,1}(u,0)\ldots 
\left(
\begin{array}{cc}
-e^{-u} + e^{u} Z_a &  2\omega (Z_a-1)S_a^-  \\
 S_a^+ (1 - Z_a) &  -\omega e^{u} + e^{-u} Z_a
\end{array} 
\right)_j 
 \ldots  \mathcal{L}_{a,L}(u,0)) 
\,.
\ee 
Comparing with the calculations of Sec. \ref{subapp:proofDG1}, in particular with eq.~\eqref{eq:toDG1}, yields the identity 
\be 
\partial_w \bm T_{\rm n}(u,w)|_{w=0}  
= - \partial_s \bm T_{\rm c}(u,0,s)|_{s=0}  
+ \frac{1}{16}[ \Ao ,[ \Ao , \partial_s \bm T_{\rm c}(u,0,s)|_{s=0}  ]]
\ee 
Plugging in the expansion of $\partial_s \bm T_{\rm c}(u,0,s)|_{s=0}$ in terms of Onsager generators yields an expansion for $\mathcal{Q}'(u)$. 
Integrating, we find
\be 
\mathcal{Q}(u) = -  \left(\sum_{m\geq 1} \frac{\tanh(N u)^{m}}{m} (\mathbf{A}_m+\mathbf{A}_{-m})  +  \tanh(N u)^2 \Ao  \right) \,.
\ee 

\section{The duality operators}
\label{app:dualityoperators}

The duality operators $D_\pm$ are defined on the lattice as limits of the $\tau_2$ transfer matrix, see Eqs.~\eqref{eq:defdualityoperators}, \eqref{eq:defdualityLaxoperators} below. In this Appendix we prove that they realize the duality automorphism of the Onsager algebra and study their fusion rules, relating them to those of the fusion category $\mathrm{TY}(\mathbb{Z}_N)$.

\subsection{Reminders on the fusion category \texorpdfstring{$\mathrm{TY}(\mathbb{Z}_N)$}{TY(ZN)}}
\label{subapp:TYZN}

In this section, we briefly introduce the basic properties of the fusion category $\mathrm{TY}(\mathbb{Z}_N)$, which we argue to be related to the duality operators.
The Tambara–Yamagami category over the finite abelian group $\mathbb{Z}_N$ \cite{Tambara:1998vmj}, denoted as $\mathrm{TY}(\mathbb{Z}_N, \chi , \epsilon)$, is a fusion category~\cite{Tensor_categories} equipped with group $G = \mathbb{Z}_N$, a nondegenerate bicharacter $\chi$ and the Frobenius--Schur indicator $\epsilon = \pm 1$. The nondegenerate bicharacter $\chi$ is a map
\begin{equation}
    \chi \; : \; G \times G \, \rightarrow \, U(1) \; ,
\end{equation}
satisfying 
\begin{equation}
    \chi (gh,k) = \chi(g,k) \chi(h,k) \; , \quad \chi (g,hk) = \chi(g,k) \chi(h,k) \; , \quad g ,h,k \in G \; .
\end{equation}

The bicharacter $\chi$ and Frobenius--Schur indicator $\epsilon$ determine the F-symbol (associator) of the category $\mathrm{TY}(\mathbb{Z}_N, \chi , \epsilon)$. In the scope of the Letter, we focus only on the fusion ring (algebra) of the category. We encourage the interested readers to consult Ref.~\cite{Tensor_categories}. We abbreviate the category $\mathrm{TY}(\mathbb{Z}_N, \chi , \epsilon)$ as $\mathrm{TY}(\mathbb{Z}_N)$ for a choice of $\chi (g,h) = \varepsilon^{2gh}$ and $\epsilon=1$. The latter must not be confused with $\varepsilon^2 = e^{\frac{2\rmi \pi}{N}}$ is the principal $N$-th root of unity.

The simple objects of the category $\mathrm{TY}(\mathbb{Z}_N, \chi , \epsilon)$ are
\begin{equation}
    \{ 0, 1, 2, \dots , N-1 , \alpha \} \; ,
\end{equation}
where $0, 1, 2, \dots , N-1 \in G = \mathbb{Z}_N$, and $\alpha$ is the non-Abelian object with quantum dimension $\sqrt{N}$.

The fusion ring (algebra) of the category $\mathrm{TY}(\mathbb{Z}_N, \chi , \epsilon)$ is given as
\begin{equation}
    x \otimes y = (x+y) \, \mathrm{mod} \, N \; , \quad x \otimes \alpha = \alpha \otimes x = \alpha \; , \quad \alpha \otimes \alpha = \bigoplus_{x=0}^{N-1} x \; ,
    \label{eq:TYZNrules}
\end{equation}
where $x ,y \in G = \mathbb{Z}_N$.

As we shall show in the following section, the duality operator (of the Onsager algebra) can be seen as a representation of the non-Abelian object $\alpha$ up to normalization.

\subsection{The duality relations}

The duality operators in the main text are defined as certain limits of the cyclic/$\tau_2$ transfer matrix,
\begin{equation}
      D_{\pm} = \lim_{u\to \mp \infty}  \varepsilon^{\frac{L}{2} \mp S^z} e^{ \pm u\, L} \bm T_{\rm c} (u,u,0) = \mathrm{Tr}_{a} \left( \prod_{j=1}^L \mathcal{L}^{\pm}_{a,j}  \right) \; .
    \label{eq:defdualityoperators}
\end{equation}
where we recall the definition  $\varepsilon=e^{\mathrm{i}\frac{\pi}{N}}$. The associated L--operators $\mathcal{L}_{a,j}^\pm$ are 
\be 
\mathcal{L}^{+}_{a,j} = 
\begin{pmatrix}
        1 &  \omega X_{a}^{-1} \\ -\varepsilon X_{a} Z_{a} & \varepsilon Z_{a} 
    \end{pmatrix}_j 
\,, \qquad
\mathcal{L}^{-}_{a,j} =
 \begin{pmatrix}
        \varepsilon Z_{a} & -  \varepsilon X_{a}^{-1} Z_{a} \\ X_{a} & \omega  
    \end{pmatrix}_j    \,.  
    \label{eq:defdualityLaxoperators}
\ee 
In order to prove that these realize a duality of the Onsager algebra, we start from the intertwining relation \eqref{eq:intertwinerK}, with $s'=w=0$ : 
\begin{equation}
    \mathsf{K}_{a,b}(s)~\big\{t,s,0\}_a \big\{ 0,0,0 \big\}_b =  
    \big\{t,0,0\}_a \big\{ 0,s,0 \big\}_b 
  \mathsf{K}_{a,b} (s) \; .
\end{equation}
Taking the limit $t\to \pm \infty$ and noticing that 
\be 
\lim_{t\to \pm\infty} e^{\mp t} \{t,s,0\}_a = e^{\mp s \sigma_j^z} \varepsilon^{- \frac{1\pm \sigma_j^z}{2}}
\mathcal{L}_{a,j}^{\mp} \,,
\ee 
we obtain 
\be 
 \mathsf{K}_{a,b}(s)~ e^{\mp s \sigma_j^z} \mathcal{L}_{a,j}^{\mp} \mathcal{L}(0,0,0)_{b,j} =  
   \mathcal{L}_{a,j}^{\mp}   \mathcal{L}(0,s,0)_{b,j} 
  \mathsf{K}_{a,b} (s)
\ee 
Using the train argument and the centrality of $\bm  T_{\rm c}(0,0,0)$, we obtain the first set of duality relations
\begin{equation}
    D_\pm  \bm T_{\rm c} (0,0,s)  = e^{\pm 2s S^z}  D_\pm  \bm T_{\rm c}(0,0,0)  \;.
\label{eq:firstdualityrelation}
\end{equation}
The second set of relations can be proved similarly: observing that $\{\widetilde{Z}_a,\widetilde{X}_a\} = \{Z_a , - \varepsilon X_a Z_a^{-1}\}$ satisfy the same algebra as $Z_a,X_a$, we have 
\be 
\lim_{w\to \pm\infty} e^{\mp w} \{0,s,w\}_a \simeq   \mathcal{L}_{a,j}^{\mp} \varepsilon^{- \frac{1\pm \sigma_j^z}{2}} e^{\mp s \sigma_j^z}
 \,,
\ee 
where $\simeq$ here stands for a unitary change of basis in the auxiliary space $a$, implementing the change $\{{Z}_a,
{X}_a\} \to \{\widetilde{Z}_a,\widetilde{X}_a\}$. Setting now $s'=t=0$  in the intertwining relation \eqref{eq:intertwinerK} and taking the limits $w\to \pm \infty$ yields 
\begin{equation}
    \bm T_{\rm c} (0,0,s) D_\pm = D_\pm e^{\pm 2s S^z} \bm  T_{\rm c}(0,0,0) \; .
\label{eq:seconddualityrelation}
\end{equation}
Inserting in \eqref{eq:firstdualityrelation}, \eqref{eq:seconddualityrelation} the expansion of $\bm T_{\rm c}(0,0,s)$ in terms of the Onsager generator $\Ai$ recovers respectively the duality relations $D_\pm\Ai = \pm \Ao D_\pm$, $\Ai D_\pm = \pm  D_\pm \Ao$ given in the main text.

\subsection{\texorpdfstring{$\mathbb{Z}_N$}{ZN} symmetry and the duality operators}
\label{subapp:ZNandduality}

We start by defining the $\mathbb{Z}_N$ symmetry of the model, 
a subgroup of the $U(1)$ symmetry. It can be generated by either choice of
\begin{equation}
    \mathsf{Z}_+   = \prod_{j=1}^{L} \varepsilon^{1-\sigma^z_j}= \prod_{j=1}^L \begin{pmatrix}
        1 & {\color{gray} 0} \\ {\color{gray} 0} & \varepsilon^2  
    \end{pmatrix}_j
    \; , \qquad
    \mathsf{Z}_-   = \prod_{j=1}^{L} \varepsilon^{1+\sigma^z_j}= \prod_{j=1}^L \begin{pmatrix}
        \varepsilon^2 & {\color{gray} 0} \\ {\color{gray} 0} & 1  
    \end{pmatrix}_j
    \; , 
\end{equation}
which are related by the $\mathbb{Z}_2$ spin-flip symmetry.
In this paragraph, we show that the operators $D_\pm$ obey 
\begin{equation}
    \mathsf{Z}_\pm \cdot D_{\pm} = D_{\pm} \cdot \mathsf{Z}_\pm = D_{\pm}  \; ,
    \label{eq:ZDrelations}
\end{equation}
which corresponds to the second set of relations in the $\mathrm{TY} (\mathbb{Z}_N)$ fusion rules, eq.~\eqref{eq:TYZNrules}.
These relations follow from local relations obeyed by the Lax operators $\mathcal{L}_{a,j}^\pm$, namely
\begin{align}
 &\varepsilon^{1\mp \sigma_j^z} \cdot \mathcal{L}_{aj}^{\pm}  
 = (V_a)^{-1} \cdot \mathcal{L}_{aj}^{\pm}  \cdot  V_a
\,, \qquad\qquad\qquad
  \mathcal{L}_{aj}^{\pm} \cdot \varepsilon^{1\mp \sigma_j^z}  
 = (W_a^{\pm})^{-1}  \cdot \mathcal{L}_{aj}^{\pm} \cdot  W_a^{\pm} \; 
\label{eq:localrelationsZn}
\end{align}
where 
\be 
V_a = (X_a)^{1/M}   \,, 
 \qquad W_a^\pm =  (X_a)^{1/M} (Z_a)^{\pm 1/M} 
 \,,
 \qquad
 q=\exp \left( \frac{\rmi \pi M}{N} \right) = \varepsilon^M
 \,. 
\ee
The local relations  \eqref{eq:localrelationsZn}, in turn, can be derived from elementary relations obeyed by the generators in the auxiliary space, $(Z_a)^{1/M} X_a (Z_a)^{-1/M} = \varepsilon^{2} X_a$ and $(X_a)^{-1/M} Z_a (X_a)^{1/M} = \varepsilon^{2} Z_a$. Diagrammatically, they can be represented as 
\begin{equation}
\begin{tikzpicture}[baseline=-1mm,scale=1.]
\draw[line width=1,blue, rounded  corners=10pt,-<,>=latex] (-1,0) -- (1,0);
\draw[line width=1,>-,>=latex] (0,-1) -- (0,1.5);
\draw[gray,fill=gray!10,thick]  (0,0) circle [radius=0.5]; \node at (0,0) {$\mathcal{L}_{a,j}^\pm$};
\draw[white,fill=white,thick]  (0,1.) circle [radius=0.25]; \node at (0,1.) {  $\varepsilon^{1\mp \sigma_j^z}$};
\end{tikzpicture}
~=~
\begin{tikzpicture}[baseline=-1mm,scale=1.]
\draw[line width=1,blue, rounded  corners=10pt,-<,>=latex] (-2,0) -- (2,0);
\draw[line width=1,>-,>=latex] (0,-1) -- (0,1);
\draw[gray,fill=gray!10,thick]  (0,0) circle [radius=0.5]; \node at (0,0) {$\mathcal{L}_{a,j}^\pm$};
\draw[blue,fill=white,thick]  (-1.25,0) circle [radius=0.4]; \node at (-1.25,0) {\small  $V_a^{-1}$};
\draw[blue,fill=white,thick]  (1.25,0) circle [radius=0.4]; \node at (1.25,0) {\small $V_a$};
\end{tikzpicture}
\qquad
\qquad
\begin{tikzpicture}[baseline=-1mm,scale=1.]
\draw[line width=1,blue, rounded  corners=10pt,-<,>=latex] (-1,0) -- (1,0);
\draw[line width=1,>-,>=latex] (0,-1.5) -- (0,1);
\draw[gray,fill=gray!10,thick]  (0,0) circle [radius=0.5]; \node at (0,0) {$\mathcal{L}_{a,j}^\pm$};
\draw[white,fill=white,thick]  (0,-1.) circle [radius=0.25]; \node at (0,-1.) {  $\varepsilon^{1\mp \sigma_j^z}$};
\end{tikzpicture}
~=~
\begin{tikzpicture}[baseline=-1mm,scale=1.]
\draw[line width=1,blue, rounded  corners=10pt,-<,>=latex] (-2,0) -- (2,0);
\draw[line width=1,>-,>=latex] (0,-1) -- (0,1);
\draw[gray,fill=gray!10,thick]  (0,0) circle [radius=0.5]; \node at (0,0) {$\mathcal{L}_{a,j}^\pm$};
\draw[blue,fill=white,thick]  (-1.25,0) circle [radius=0.4]; \node at (-1.25,0) {\small  $W_a^{-1}$};
\draw[blue,fill=white,thick]  (1.25,0) circle [radius=0.4]; \node at (1.25,0) {\small $W_a$};
\end{tikzpicture}
\end{equation}
Using these relations repeatedly indeed recovers the global relations \eqref{eq:ZDrelations}.

\subsection{Fusion of two duality operators}
\label{subapp:dualityfusion}

In this section we prove the last set of fusion rules, associated with the fusion of two duality operators.
Start with the product 
\begin{equation}
\begin{split}
    D_{-} \cdot D_{-} & = \mathrm{Tr}_{a ,b} \left( \prod_{j=1}^L \mathcal{L}_{aj}^{-}\mathcal{L}_{bj}^{-} \right) \\ & = \mathrm{Tr}_{a , b} \left( \prod_{j=1}^L \begin{pmatrix}
        \varepsilon^2 Z_{a} Z_{b} - \varepsilon \omega Z_{a} X_{a}^{-1} X_{b} &  -\varepsilon^2 \omega Z_{a} Z_{b}  X_{b}^{-1} - \varepsilon \omega^2 Z_{a} X_{a}^{-1}  \\ \varepsilon Z_{b} X_{a} + \omega X_{b} & - \varepsilon \omega Z_{b} X_{a} X_{b}^{-1} + \omega^2
    \end{pmatrix}_j \right) \; .
\end{split}
\end{equation}
Noting that all entries in the product of Lax operators $\mathcal{L}_{aj}^{-}\mathcal{L}_{bj}^{-}$ commute with $\mathcal{K}=Z_{a} X_{a}^{-1} X_b$, we can decompose the doubled auxiliary space as $\bigoplus_{k=0}^{N-1} k$, where $k$ denotes the $N$-dimensional space where $\mathcal{K}$ has eigenvalue $\mu_k=\varepsilon^{N-1} \omega^k= -\varepsilon \omega^k$, namely 
\be 
 D_{-} \cdot D_{-} = \sum_{k=0}^{N-1} \mathrm{Tr}_{k}
 \prod_{j=1}^L  \left. (\mathcal{L}_{aj}^{-}\mathcal{L}_{bj}^{-})  \right|_{k} \,,
\ee 
where $(\mathcal{L}_{aj}^{-}\mathcal{L}_{bj}^{-})|_{k}$ denotes the projection of the doubled Lax operator onto the auxiliary subspace with eigenvalue $\mu_k$.
The form of the eigenvalues $\mu_k$ can be deduced from the observation that $(\mathcal{K})^N=\omega^{\frac{N(N-1)}{2}}=\varepsilon^{N(N-1)}$. Within a $k$-subspace, the action of all remaining operators can be recast in terms of a pair of new operators, $Z_k=\varepsilon^{-2}\omega^{-k-1} Z_a Z_b$, $X_k=X_b$, satisfying the usual $\mathbb{Z}_N$-clock algebra $Z_k X_k = \omega X_k Z_k$, $Z_k^N = X_k^N = 1$. We then have   
\be 
\left. (\mathcal{L}_{aj}^{-}\mathcal{L}_{bj}^{-})  \right|_{k}
   =
    \begin{pmatrix}
    \omega^{k+1} (1+Z_k) &  \omega^{k+2} (1-Z_k )X_{k}^{-1}  \\\omega X_{k}(1-Z_k) & \omega (Z_k + \omega)
    \end{pmatrix}_j
  = 
      \begin{pmatrix}
    \omega^{k+1}  &  0  \\  0  & \omega
    \end{pmatrix}_j 
    \cdot
    \mathcal{L}_{kj} (0,0,0) = \begin{pmatrix}
        1 + Z_k & \omega(1-Z_k) X_k^{-1} \\ X_k (1-Z_k) & \omega + Z_k
    \end{pmatrix}_j  \,.
\ee 
Recognizing the rightmost matrix in the above formula as $\mathcal{L}_{k,j}(0,0,0)$, we can therefore factorize the product $D_- \cdot D_-$ as 
\be 
 D_{-} \cdot D_{-} = \left( \sum_{k=0}^{N-1} 
 \prod_{j=1}^L      \begin{pmatrix}
    \omega^{k+1}  &  0  \\  0  & \omega
    \end{pmatrix}_j    \right)  T(0)
=  \omega^L 
\left(\sum_{k=0}^{N-1} \omega^{k\left(S^z+\tfrac{L}{2}\right)}\right)
~ T(0)
\ee 
hence proportional to the projector onto the sector with magnetization $S^z = -\tfrac{L}{2} \,~ \mathrm{mod} \, N$. 
Noticing that this corresponds to the sector where $\mathsf{Z}_-=1$, we can also write 
\be 
 D_{-} \cdot D_{-}  = \omega^L ~\left( \sum_{k=0}^{N-1} (\mathsf{Z}_-)^k \right) ~  T(0) \,.
\ee 
When $T(0)$ is invertible, we recover an explicit representation of the fusion algebra of the $\mathrm{TY}(\mathbb{Z}_N)$ category.
Analogously, we can prove the following relation for the fusion of $D_\pm$, 
\be 
 D_{+} \cdot D_{+}  = \left( \sum_{k=0}^{N-1} (\mathsf{Z}_+)^k \right) ~  T(0) \,.
\ee 

\end{document}